\documentclass[%
 amsmath,amssymb,
 twocolumn,
]{revtex4-2}

\usepackage{graphicx}
\usepackage[caption=false]{subfig}
\usepackage{dcolumn}
\usepackage{bm}
\usepackage{braket}
\usepackage{tikz}
\usetikzlibrary{decorations.markings}
\usetikzlibrary{decorations.pathreplacing, arrows.meta}
\usetikzlibrary{positioning}
\usepackage{amsmath}
\usepackage{hyperref}

\hypersetup{
	colorlinks,
	linkcolor={red!50!black},
	citecolor={blue!80!black},
	urlcolor={blue!50!black}
}

\newcommand{\rsup}[3]{(\mathbf{#1},\mathbf{#2},#3)}
\newcommand{\rsAB}[4]{(\mathbf{#1}_{#2},\mathbf{#1}_{#3},#4)}
\newcommand{\measV}[1]{\mathrm{d}^{3} {#1}}
\newcommand{\meas}[1]{\mathrm{d}{#1}}

\newcommand{\paraG}{\mathbb{P}}
\newcommand{\perpG}{\mathbb{S}}

\begin{document}

\preprint{APS/123-QED}

\title{Atoms near a conducting wedge: decay rates and entanglement around a corner}

\author{Romuald Kilianski}
\email{r.kilianski.1@research.gla.ac.uk}
\author{Robert Bennett}%
\affiliation{School of Physics \& Astronomy, University of Glasgow, Glasgow G12 8QQ, United Kingdom}

\date{\today}

\begin{abstract}
The behavior of an atomic system is influenced by introducing a metallic surface. This work explores how the decay landscape can be altered by the presence of sharp corners. We examine two scenarios: the modified spontaneous decay of a single atom, which leads us to speculate about potential applications in microscopy, and the case of a more fundamental, theoretical interest --- the behavior of an entangled pair of atoms near a corner. The latter, when two atoms are positioned ``out of the line of sight," opens up a possible line of investigation into devices which are able to ``see around corners''.
\end{abstract}
\maketitle

\section{\label{sec:level1}Introduction}
The science of light emission is fundamental to quantum optics and photonics, profoundly impacting modern technology and everyday experiences. At its heart lies spontaneous emission, once believed to be an intrinsic atomic property \cite{haar1967old}. However, since the development of quantum electrodynamics, it can be understood as a quantum-mechanical phenomenon arising from the fact that the atomic energy eigenstates are not eigenstates of the combined matter-field system of the atom coupled to light. 

Spontaneous emission can be manipulated by directly interacting with the emitter, e.g., via the Stark shift or by modifying its immediate environment. The term ``spontaneous'' hides the fact that the dipole interacts with the electromagnetic vacuum field \cite{milonni2013quantum}, and any change to boundary conditions alters the available mode structure. Hence, the atom's environment affects its light-emitting properties, including, for example, the radiative decay rate 
\cite{purcell1995spontaneous}.
In the Wigner-Weisskopf approximation \cite{novotny2006principles,Weisskopf1997}, the radiative decay rate can be calculated using Fermi's Golden rule \cite{loudon2000quantum}. In such a calculation, one encounters an associated quantity called the local density of states (LDOS)\cite{LDOS_Joulain}. This counts how many excitation modes are available at a location in space (per unit volume), which can also be understood as the density of vacuum fluctuations at that point. The emission rate, seen from this perspective, can be understood to be proportional to the electric field that the atom (dipole) produces at a given location \cite{novotny2006principles}. This dependence of radiative properties on the spatial arrangement motivates the pursuit of engineered geometries capable of influencing the atomic response (for a review, see \cite{Pelton2015}) to produce enhancement or suppression of the spontaneous decay rate. 

\begin{figure}
        \includegraphics[width=\columnwidth]{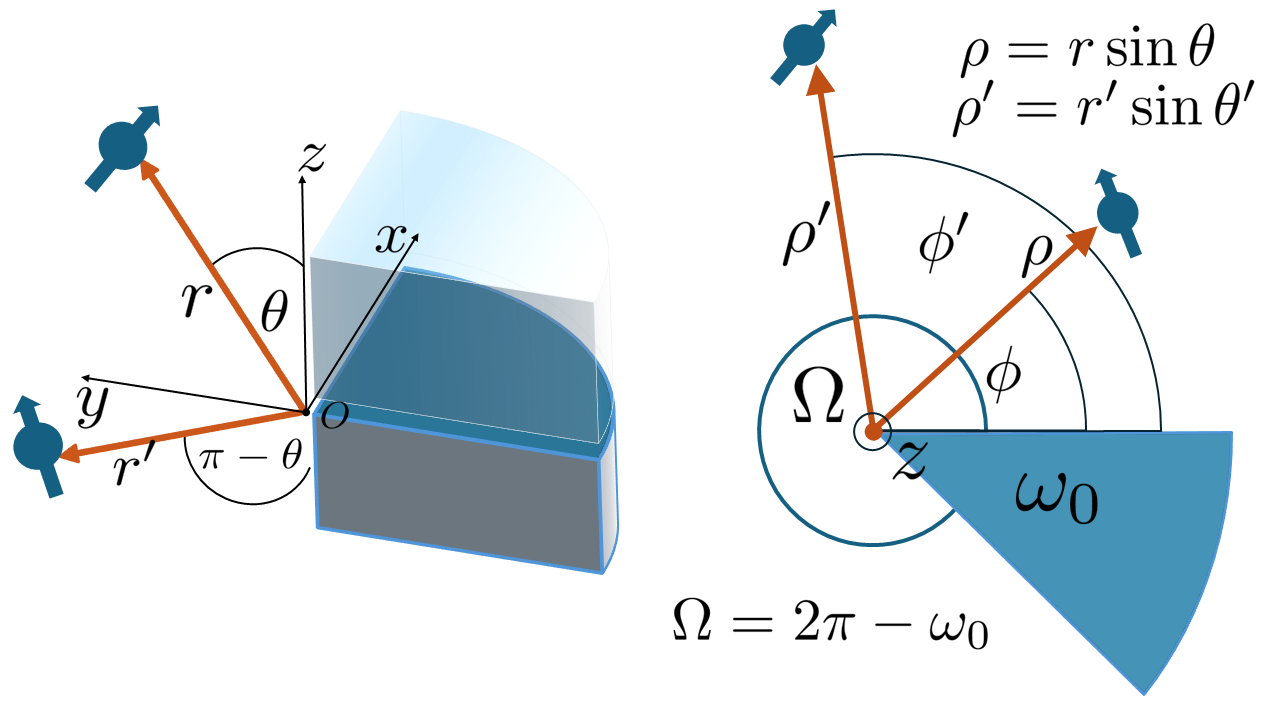}
    \caption{Illustration of the three coordinate systems used to describe the wedge of internal angle $\omega_{0}$ and the external angle $\Omega$. The edge of the wedge coincides with the $z$ axis, and the origin $O$, marks the location of its tip.}
    \label{fig:wedge_coords}
\end{figure}

The first modification scheme was introduced in a remarkably short note by Purcell \cite{purcell1995spontaneous}, in which it was predicted that magnetic transition rates could be enhanced by confining an atom to a small volume. The quantum yield must be considered to measure the photons emitted via spontaneous decay in this way. This is the proportion of the time that the relaxation of the atom coincides with photon emission. Purcell's original proposal was an example of a cavity-based mechanism \cite{HarocheQED}, which requires a balance between the cavity dimensions and the storage duration, both of which are parameters that affect the quantum yield. The presence of the cavity can increase the quantum yield only if the initial quantum yield is small, such as in dye molecules, and decrease it only in high-yield, modern emitters like quantum dots, \cite{berezovsky2008QDot,press2008QDot}. As a result of this limitation, the focus has shifted to more exotic structures, such as photonic crystals (PCs)\cite{joannopoulos1997photonic,Cersonsky2021}, and plasmonic metal nanostructures (PMCs) \cite{Pelton2015,NovotnyPMC,KuhnPMC,CARMINATIPMC}. PCs are optical structures that can restrict light propagation within specific frequency ranges \cite{BaduguPC,wangPC2002,HatefPC,ogawaPCl,aokiPC2008}. This results in a so-called photonic bandgap, arising due to multiple interference events within the crystal, which leads to an increase in LDOS \cite{LDOSBG}.
PMCs such as metallic nanoparticles support plasmon resonances \cite{forestiere2020quantum} - oscillations of electric charges on the material's surface- that couple to the electric field modes, thereby strongly modifying the LDOS. Some recent works investigated different geometries, such as nanocones and nanocylinders \cite{ConesCyls23}, as well as layered nanostructures \cite{Pors2015}.

We will also study a process closely related to the spontaneous decay rate, namely the \textit{cooperative} decay rate (CDR). This results from an interplay between two or more atoms in an ensemble, which gains environment dependence like the spontaneous decay rate. The coupled interaction is borne out of the synchronisation of the dipole moments, resulting in significant modification of their emission properties. The cause can be understood as the excitation being reabsorbed and then emitted multiple times between the emitters before the photon is released \cite{FriedbergCDR,ManassahCDR}.  Dicke first introduced the phenomenon; in his seminal work \cite{Dicke}, he studied cooperative radiative effects in an ensemble of atoms. From then, a vast array of literature emerged covering superradiance,  \cite{arecchi1970cooperative,bolda1995superfluorescence,feher1958spontaneous,malcuit1987transition} - an enhancement of the decay rate, as well as subradiance \cite{crubellier1987superradiance,stroud1972superradiant} - suppression of the decay rate. Some examples of the recent works studying CDR explored it in the context of qubits \cite{Mlynek2014,RastogiQubit}, trapping \cite{MassonTrap}, and a photonic quantum engine \cite{Kim2022}.

In this paper, we investigate both the decay rate and the CDR processes that are modified by introducing a metallic structure -- a review showing analytical solutions for simple geometries can be found in Ref.~\cite{ManassahGeom}.
Here, we model the atom's environment as perfectly reflecting surfaces embedded in a vacuum.  We investigate how radiative interactions are altered by placing the emitter(s) near a metallic wedge of arbitrary opening angle. The wedge geometry interests us for at least two reasons:  we envisage an interest in the possibility of controlling and sensing changes to the decay landscape from an obscured, ``around the corner" point of view. For the latter --- the particular case of the CDR --- we study the simplest ensemble (consisting of two atoms), producing super- and sub-radiant effects -- a model (without the surface), experimentally realised in \cite{DeVoeTwoAtoms, eschner2001TwoAtoms}.  

We will begin, however, by investigating single-atom decay rates with a possible long-term application to the microscopy of biological samples.  A first step in imaging such a sample is often labelling it by populating it with fluorescing molecules (dye). Typically, this means the sample will end up containing sources of illumination that are spaced closer than the diffraction limit, providing a barrier to high-resolution imaging. Several approaches have been developed to circumvent this limitation \cite{DiffLimHuang2010}, among them, the stimulated emission depletion (STED) technique \cite{STEDReview, STEDGaliani2012, STEDHell}. It achieves super-resolution by ``switching off'' the edge of the luminescent region, effectively producing a focal spot with a smaller diameter. This is achieved by illuminating the area of interest with a doughnut-shaped laser beam with zero intensity at its centre. This results in distinct decay rates at the beam's centre and periphery, allowing for better resolution of the sample's features. One could envisage replacing the doughnut-shaped laser beam with a metallic ring or hollow cylinder, which would effectively take the role of the laser in providing a differential decay rate to the dye molecules. The use of metallic objects in this context shares some features with known techniques of metal-enhanced fluorescence \cite{MEFgeddes2002, MEFReview}, exploited in, e.g. F\"{o}rster resonance energy transfer, where the plasmon coupling plays a significant role \cite{FRETPoirier2015l}.

The direct modification of LDOS in microscopy also exists in tip-enhanced near-field optical microscopy (TENOM) \cite{SharpEdgeHartschuh2014}, where the presence of a sharp structure such as an antenna facilitates the two-way conversion of radiation into localised spots of energy. The-wavelength-regime-appropriate optical antennas have been of considerable interest in nanophotonics, re-purposing the well-established radio- and microwave-antenna technology to enhance and suppress atomic processes in the nanometre regime \cite{OpticalAntBharadwaj, OpticalAntNovotny2011, CambridgeOpticalAnt}.   
The most pertinent region that can display similarities with the case of a sharply-tipped antenna is, of course, near the corner constituting the ring's inner edge. This is also the type of geometry for which decay rate results are not yet present in the literature. Thus, in our idealised, perfectly reflecting model, we focus only on the influence of the local (corner or wedge) geometry on the radiative decay rates. 

This paper is organised as follows: in Section \ref{sec2}, we review Green's tensor formalism that allows for the computation of decay rates in complex geometries and its connection to the Hertz vector potentials (both of which are representations of electromagnetic field propagation between source and observation points). In Section \ref{sec3}, we describe the setup of our geometry and delineate the different forms of the sought solutions. Following that, in Section \ref{sec4}, we obtain summation formulations from the two different methods of calculating the electric fields scattered by wedges, allowing us to perform controlled and convergent numerical computations. Section \ref{sec5} presents some example results of the presented formulae in contour plots of the landscape of decay rates before describing potential applications in more detail.

\section{Fields as an impulse response}
\label{sec2}

The spontaneous decay rate $\Gamma$ and the CDR $\Gamma^{C}$ are quantities that can be obtained from the imaginary part of the electromagnetic (dyadic) Green's tensor $\mathbb{G}(\mathbf{r},\mathbf{r}',\omega)$ (see, e.g., \cite{novotny2006principles,tai1994dyadic,chew1995waves}), capturing the linear relationship between the observed electric field at position $\mathbf{r}$ and the source at position $\mathbf{r}'$, at a frequency $\omega$. In a region of relative permittivity $\varepsilon \equiv \varepsilon(\mathbf{r},\omega)$ and unit relative permeability, it is a solution to the inhomogeneous Helmholtz equation
\begin{align}
\nabla\times\nabla\times \mathbb{G}(\mathbf{r},\mathbf{r}',\omega) - k^{2}\mathbb{G}(\mathbf{r},\mathbf{r}',\omega) = \mathbb{I}\delta(\mathbf{r}-\mathbf{r}'),
\end{align}
where $k^2 = \varepsilon\omega^2/c^2$ ($c$ being the speed of light), $\mathbb{I}$ is the unit dyadic (i.e., the $3\times 3$ identity matrix) and $\delta$ is the Dirac delta function. Solutions to this equation may be written as
\begin{align}\label{Gandg}
\mathbb{G}\rsup{r}{r'}{\omega} = \left(\mathbb{I}+\frac{1}{k^{2}}\nabla\otimes\nabla\right)g(\mathbf{r},\mathbf{r'},\omega),
\end{align}
where $g(\mathbf{r},\mathbf{r}',\omega)$ is a frequency-dependent scalar Green's function satisfying
\begin{align}
 (\nabla^{2} + k^{2})g(\mathbf{r},\mathbf{r}',\omega) = -\delta(\mathbf{r}-\mathbf{r}').   
\end{align}
The appeal of using $\mathbb{G}(\mathbf{r},\mathbf{r}',\omega)$ lies in its ability to encapsulate the complete electromagnetic response of the system. It contains information that can be used to learn about many physical processes, e.g. the Lamb shift \cite{FriedbergCDR}, and resonance energy transfer \cite{FRET}, among others; for a recent experimental account verifying a range of quantities obtained from $\mathbb{G}(\mathbf{r},\mathbf{r}',\omega)$, see Ref.~\cite{Rustomji}. 

In the search for the geometry-specific $\mathbb{G}(\mathbf{r},\mathbf{r}',\omega)$ for our wedge, it will be helpful to derive an alternative representation of the source-observer relationship, developed with the help of vector fields, using so-called Hertz (vector) potentials \cite{Zangwill_2012}. This approach is favoured in engineering and often found in problems describing, e.g. antenna radiation \cite{galejs1969antennas}. Here, we use it to help us compute specific components of $\mathbb{G}(\mathbf{r},\mathbf{r}',\omega)$.    

We can describe the response to a time-harmonic infinitesimal current distribution $\mathbf{j}(\mathbf{r})$ via the vector Hertz potential $\overline{\bm \Pi}(\mathbf{r})$:
\begin{align}
\overline{\bm{\Pi}}(\mathbf{r}) = \frac{i Z}{k}\int \measV{\mathbf{r}'}\: g(\mathbf{r},\mathbf{r'},\omega)\mathbf{j}(\mathbf{r}'),
\label{Hertz}
\end{align}
where $Z = \mu_0\omega/k$ is the impedance (with $\mu_0$ the permeability of free space).
The potential can then be used to express the electric field as\footnote{or equivalently as \begin{equation}
\mathbf{E}(\mathbf{r}) = (k^{2} + \nabla\otimes \nabla\cdot) \overline{\bm{\Pi}}(\mathbf{r}).\\
\end{equation}}
\begin{equation}
\label{EFieldInTermsOfHertz}
\mathbf{E}(\mathbf{r}) = k^{2}  \overline{\bm{\Pi}}(\mathbf{r}) + \nabla[\nabla\cdot \overline{\bm{\Pi}}(\mathbf{r})].
\end{equation}

As an aside we note that one can find the electric field $\mathbf{E}^{w}$ of a current directed along an arbitrary unit vector $\hat{\mathbf{w}}$ by substituting ${\mathbf{j}(\mathbf{r}) = \hat{\mathbf{w}}\delta(\mathbf{r}-\mathbf{r}')/(i k Z)}$ into \eqref{EFieldInTermsOfHertz} via \eqref{Hertz}, then simplifying using \eqref{Gandg} to find:
\begin{align}\label{electricFieldInDirectionw}
\mathbf{E}^{w}(\mathbf{r}) = \mathbb{G}(\mathbf{r},\mathbf{r}',\omega)\cdot \hat{\mathbf{w}}.
\end{align}
While we are not directly dealing with electric fields in this paper, we are, however, considering the Green's tensor $\mathbb{G}$ in detail, which can then be used to calculate electric fields via \eqref{electricFieldInDirectionw} for the setup described here. 

To calculate the quantities of interest, we use the perturbative methods of macroscopic QED \cite{buhmann2013dispersion}. The electromagnetic response contained in $\mathbb{G}(\mathbf{r},\mathbf{r}',\omega)$ is fully classical at the level of the permittivity but finds its place as amplitude in the expression of the quantised electromagnetic field in the presence of dispersive and absorbing dielectric media (see, e.g., \cite{RobNicMQED,novotny2006principles}),
\begin{align}\label{EFieldDef}
\nonumber
\hat{\mathbf{E}}(\mathbf{r},t) = \int_{0}^{\infty}\!&\meas{\omega}\sum_{\lambda=e,m}\int\measV{\mathbf{r}'}\bigg[ \mathbb{G}_{\lambda}(\mathbf{r},\mathbf{r}',\omega)\cdot \hat{\mathbf{f}}_{\lambda}(\mathbf{r}',\omega)e^{-\mathrm{i}\omega t}\\
&+\mathbb{G}^{*}_{\lambda}(\mathbf{r},\mathbf{r}',\omega)\cdot \hat{\mathbf{f}}^{\dagger}_{\lambda}(\mathbf{r}',\omega)e^{\mathrm{i}\omega t} \bigg],
\end{align}
alongside $\hat{\mathbf{f}}_{\lambda}(\mathbf{r},\omega)$ and $\hat{\mathbf{f}}_{\lambda}^{\dagger}(\mathbf{r},\omega)$ which are annihilation, and creation operators, respectively, satisfying a bosonic commutation relation
\begin{align}\label{commutator}
\left[ \hat{\mathbf{f}}_{\lambda}(\mathbf{r},\omega), \hat{\mathbf{f}}_{\lambda'}^{\dagger}(\mathbf{r}',\omega')\right] = \delta_{\lambda \lambda'}\delta(\omega-\omega')\mathbf{\delta}(\mathbf{r}-\mathbf{r}').
\end{align}
and with electric and magnetic components of the Green's tensor given by
\begin{align}
\mathbb{G}_{e}(\mathbf{r},\mathbf{r}',\omega) =&\, i \frac{\omega^{2}}{c^{2}}\sqrt{\frac{\hbar}{\pi \varepsilon_{0}} \mathrm{Im}\varepsilon(\mathbf{r}',\omega)}\mathbb{G}(\mathbf{r},\mathbf{r}',\omega),\\
\mathbb{G}_{m}(\mathbf{r},\mathbf{r}',\omega) =& \,i \frac{\omega}{c}\sqrt{\frac{\hbar }{\pi \varepsilon_{0}}\frac{\mathrm{Im}\mu(\mathbf{r}',\omega)}{|\mu(\mathbf{r}',\omega)|^{2}}}\left[\nabla'\times\mathbb{G}(\mathbf{r},\mathbf{r}',\omega)\right]^{\mathrm{T}},
\end{align}
where $^{\mathrm{T}}$ denotes the transpose and $\varepsilon_{0}$ is the permittivity of free space.

We begin by focusing on the atomic decay rate $\Gamma$, an inverse of the atomic lifetime $\tau$. This can be calculated using Fermi's Golden rule, where we denote the initial $i$ and final $f$ states, and $E_{f}$ and $E_{i}$ are their corresponding energies,
\begin{align}
\Gamma = \frac{2\pi}{\hbar}\sum_{i}|\bra{f}\hat{H}_\mathrm{int}\ket{i}|^{2}\delta(E_{f}-E_{i}),
\end{align}
where $\hat{H}_{\mathrm{int}}$ is the interaction Hamiltonian, $\hat{H}_{\mathrm{int}}=-\hat{\mathbf{d}}\cdot \hat{\mathbf{E}}$, with $\hat{\mathbf{d}}$ being the dipole operator.

The spontaneous decay rate (see Ref.~\cite{RobNicMQED, novotny2006principles} for derivation), of an atom at position $\mathbf{r}_{\mathrm{A}}$ with electric dipole moment $\hat{\mathbf{d}}_{\mathrm{A}}$, and transition frequency $\omega$, is given by
\begin{align}\label{decayRate}
\Gamma = \frac{2 \omega^{2}}{\hbar \varepsilon_{0}c^{2}}\mathbf{d}_{\mathrm{A}}\cdot \mathrm{Im}\mathbb{G}\rsAB{r}{\mathrm{A}}{\mathrm{A}}{\omega_\mathrm{A}}\cdot\mathbf{d}_{\mathrm{A}}.
\end{align}

The CDR can be calculated similarly. The collective behaviour of an ensemble of $N$ atoms is usually calculated via a master equation \cite{CDRMasterAgarwal1974}. Still, given the elementary nature of the ensemble $(N=2)$, it can be calculated directly using Fermi's Golden Rule \cite{PowerCDR}. Proceeding via the latter method --- as detailed in Appendix \ref{ap} --- we have
\begin{align}
\label{CDR 1}
\Gamma^{\mathrm{C}} = \frac{\Gamma_{\mathrm{A}}}{2} + \frac{\Gamma_{\mathrm{D}}}{2}\pm \Gamma_{\mathrm{dd}},
\end{align}
where for $i \in \{\mathrm{A},\mathrm{D}\}$ 
\begin{equation}
\Gamma_{i} = \frac{2 \omega^{2}}{\hbar \varepsilon_{0}c^{2}}\hat{\mathbf{d}}_{i}\cdot \mathrm{Im}\mathbb{G}\rsAB{r}{i}{i}{\omega}\cdot\hat{\mathbf{d}}_{i},
\end{equation}
and
\begin{equation}
\Gamma_{\mathrm{dd}} = 
\frac{2 \omega^{2}}{\hbar \varepsilon_{0}c^{2}}\hat{\mathbf{d}}_{\mathrm{A}}\cdot \mathrm{Im}\mathbb{G}\rsAB{r}{\mathrm{A}}{\mathrm{D}}{\omega}\cdot\hat{\mathbf{d}}_{\mathrm{D}}.
\end{equation}
Equation (\ref{CDR 1})   describes a cooperative decay rate in a system with a single photon and two atoms at positions $\mathbf{r}_\mathrm{A}$(acceptor) and $\mathbf{r}_\mathrm{D}$(donor), where we took their respective transition frequencies to be equal, i.e. $\omega \equiv\omega_{\mathrm{A}} = \omega_{\mathrm{D}}$. The first two terms are the respective probabilities of each atom decaying into the ground state, which triggers the cooperative dipole-dipole interaction $\pm\Gamma_{\mathrm{dd}}$. The sign with which $\Gamma_{\mathrm{dd}}$ contributes is controlled by the initial symmetric ($+$) or anti-symmetric ($-$) atomic entanglement between $\mathrm{A}$ and $\mathrm{D}$, ultimately determining the super-, or sub-radiant nature of the entire process.  
\section{The setup}
\label{sec3}

The setup consists of a perfectly electrically conducting wedge whose edge is infinitely long and coincides with the $z$-axis, as depicted in Fig.~\ref{fig:wedge_coords}. The atoms are represented as point-like dipoles. 

The outer wedge angle $\Omega$ is a function of the internal angle $\omega_{0}$, i.e., $\Omega = 2\pi - \omega_{0}$.

To produce a complete picture of the scattering behaviour due to an oscillating electric dipole near a perfectly reflecting wedge, one would, in principle, have to compute all the elements of the dyadic Green's tensor.  For the case of spontaneous decay, the structure of Eq.~\eqref{decayRate} means that we need only obtain the diagonal elements of $\mathbb{G}$. For the cooperative decay rate, however, we generally require the whole Green's tensor, albeit specific dipole orientations can reduce the number of elements needed --- the issue to which we will return.

The usual formulation of the Green's tensor in terms of  Fourier integrals (see, e.g. \cite{tai1994dyadic}) is especially difficult to implement for the wedge geometry since the contained integrals display highly oscillatory behaviour and consequently produce non-converging results when computed numerically. We, therefore, turn to alternative approaches, where we re-express the integrals as doubly infinite series. These techniques were initially developed for applications to problems where source-observer separation is significant or the observation point lies far away from the edge - avoiding the singular behaviour of the Green's tensor at the coincidence limit. In our case, we avoid these limitations by seeking only the imaginary part of $\mathbb{G}(\mathbf{r},\mathbf{r}',\omega)$, which stays bounded for $\mathbf{r}'\rightarrow \mathbf{r}$, allowing us to investigate the behaviour in the near-field.

We can decompose the Green's tensor into a part $\paraG$ that depends on the source dipole parallel to the edge of the wedge ($z$-axis) and a part $\perpG$ depending on the perpendicular dipole orientation, writing (using a Cartesian basis) \footnote{The Green's tensor which connects exclusively the $z$- oriented dipole to $x,y,z$ observation directions is $$\begin{pmatrix}
0 & 0 & 0\\
0 & 0 & 0\\
\mathbb{G}_{x z}& \mathbb{G}_{y z}&\mathbb{G}_{zz}
\end{pmatrix},$$  however, in reciprocal media $\mathbb{G}(\mathbf{r},\mathbf{r}',\omega) = \mathbb{G}^{\mathrm{T}}(\mathbf{r}',\mathbf{r},\omega)$, thus the remaining components in the last column of $\paraG$ can be obtained from the above. }
\begin{equation}
\mathbb{G}(\mathbf{r},\mathbf{r}',\omega) = \paraG(\mathbf{r},\mathbf{r}',\omega) + \perpG(\mathbf{r},\mathbf{r}',\omega)
\end{equation}
with 
\begin{align}
\paraG(\mathbf{r},\mathbf{r}',\omega) &= \begin{pmatrix}
0 & 0 & \mathbb{G}_{zx}\\
0 & 0 & \mathbb{G}_{zy} \\
\mathbb{G}_{x z}& \mathbb{G}_{y z}&\mathbb{G}_{zz}
\end{pmatrix} \nonumber
\end{align}
and the resulting implicit definition of $\perpG$ via;
\begin{align}
\label{G perp}
\perpG(\mathbf{r},\mathbf{r}',\omega) &= \mathbb{G}(\mathbf{r},\mathbf{r}',\omega) -\paraG(\mathbf{r},\mathbf{r}',\omega). 
\end{align}
We make this decomposition because, in the end, we will employ different methods to seek solutions for the parallel and perpendicular parts, respectively. Ultimately, this is because $\paraG(\mathbf{r},\mathbf{r}',\omega)$ and $\perpG(\mathbf{r},\mathbf{r}',\omega)$ display different behaviours concerning convergence once the inevitable truncation to a finite number of terms is made. In particular, the parallel part is $\paraG$ better suited to the cylindrical formulation, whereas a spherical wave expansion describes the perpendicular part $\perpG$  is described more accurately by a spherical wave expansion.

\section{Results}
\label{sec4}
\subsection{Green's tensor components originating from a parallel dipole}
\begin{figure}
\begin{tikzpicture}

    \definecolor{pathcolor}{rgb}{15,0,0}

    \definecolor{color1}{rgb}{0.8,0.8,0.8} 

    \fill[color1] (-2, 0) rectangle (-1, 3);
    \fill[color1] (0, 0) rectangle (1,3);
    \fill[color1] (2, 0) rectangle (3,3);
    \fill[color1] (-3,-3) rectangle (-2,0);
    \fill[color1] (-1,-3) rectangle (0,0);
    \fill[color1] (1,-3) rectangle (2,0);

    \tikzset{->-/.style={decoration={
            markings,
            mark=at position #1 with {\arrow{>}}},postaction={decorate}}}
            
     \tikzset{-<-/.style={decoration={
            markings,
            mark=at position #1 with {\arrow{<}}},postaction={decorate}}}
    
    \draw[->] (-3, 0) -- (3, 0) node[right] {$\text{Re } \alpha$};
    \draw[->] (0, -3) -- (0, 3) node[above] {$\text{Im } \alpha$};
    
    \draw[dashed] (-2, -3) -- (-2, 3) node[above] at (-2.4, 0) {$-2\pi$};
    \draw[dashed] (-1, -3) -- (-1, 3) node[above] at (-1.3, 0) {$-\pi$};
    \draw[dashed] (1, -3) -- (1, 3) node[above] at (0.8, 0) {$\pi$};
    \draw[dashed] (2, -3) -- (2, 3) node[above] at (1.75, 0) {$2\pi$};


    \draw[pathcolor, thick,decoration={markings, mark=at position 0.6 with {\arrowreversed{Stealth}}},postaction={decorate}] plot[smooth, tension=2] coordinates {(-0.5, 3) (0.5, 0.7) (1.5, 3)} node[above] at (0.5, 0.2) {$\gamma_{+}$};
    \draw[pathcolor, thick,decoration={markings, mark=at position 0.7 with {\arrow{Stealth}}},postaction={decorate}] plot[smooth, tension=2] coordinates {(-1.5, -3) (-0.5, -0.7) (0.5, -3)} node[above] at (-0.5, -0.7) {$\gamma_{-}$};
    \draw[dashed,thick, red,decoration={markings, mark=at position 0.3 with {\arrow{Stealth}}},postaction={decorate}] (0.5,-3) .. controls (0.5,0) and (1.5,0) .. (1.5,3) node[above] at (-0.5, 0.6) {$\gamma_{\mathrm{L}}$};
    \draw[dashed, thick, red, decoration={markings, mark=at position 0.3 with {\arrow{Stealth}}},postaction={decorate}] (-0.5,3) .. controls (-0.5,0) and (-1.5,0) .. (-1.5,-3) node[above] at (1.5, 0.6) {$\gamma_{\mathrm{R}}$};
    \fill[black] (0, 1.5) circle (0.1) node[above] at (0, 1.6) {$\mathrm{i} \xi$}; 
    \fill[black] (0, -1.5) circle (0.1) node[above] at (0, -2.1) {$-\mathrm{i} \xi$}; 

    \fill[black] (2, 1.5) circle (0.1) node[above] at (2, 1.6) {} ; 
    \fill[black] (2, -1.5) circle (0.1) node[above] at (2, -2.1) {};

    \fill[black] (-2, 1.5) circle (0.1) node[above] at (-2, 1.6) {} ; 
    \fill[black] (-2, -1.5) circle (0.1) node[above] at (-2, -2.1) {};

\end{tikzpicture}
\caption{Deformed integration path for the Sommerfeld-Maliuzhinets contour. The upper and lower loops (red) extend to positive and negative infinity, respectively, in the imaginary direction, where the function in the integrand tends to zero. The dashed paths (red) are in opposite directions and their contributions mutually cancel out. The points $\pm\mathrm{i}\xi\pm 2n\pi $ for $n \in \mathbb{N}$, show locations of the branch points.}
\label{complex}
\end{figure}
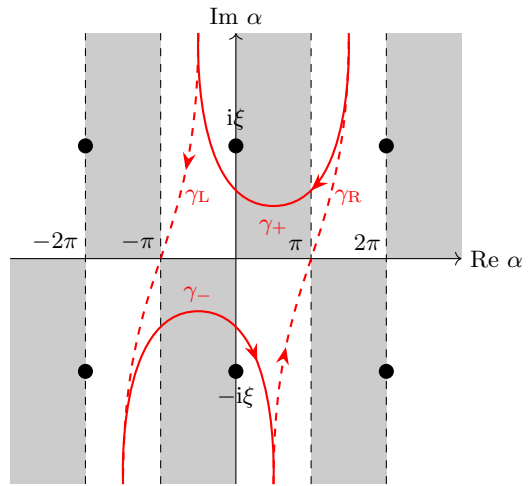

Here, we develop the tools that will allow us to obtain the imaginary part of $\paraG(\mathbf{r},\mathbf{r}',\omega)$. This will be done via the intermediate step of computing the $z$-component of the Hertz vector potential, Eq.~\eqref{Hertz}. Since the edge of the wedge runs infinitely along the $z$ axis, a cylindrical coordinate system seems like the most natural choice for this setup --- depicted in Fig. \ref{fig:wedge_coords}. If we orient the dipole moment to be parallel to the edge, $\mathbf{p} = \frac{\hat{z}}{i k Z}\delta(\mathbf{r}- \mathbf{r}')$, we obtain the Hertz vector potential that is non-zero only in its $z$ component, i.e. $\overline{\Pi}(\mathbf{r}) = \{0,0,\Pi_{z}(\mathbf{r})\}$, consequently defining our choice for $\paraG(\mathbf{r},\mathbf{r}',\omega)$. Some computational effort can be saved by selectively calculating the elements of the ``parallel'' Green's tensor; to this effect, we set up the quantity of interest as
\begin{align}
\nonumber
\mathrm{Im}\paraG_{zz}(\mathbf{r},\mathbf{r}',\omega) &=\hat{z}\cdot \mathrm{Im}\mathbb{G}\cdot \hat{z}\\
\nonumber
&= \hat{z}\cdot(k^{2} + \nabla\otimes\nabla\cdot)\mathrm{Im}\overline{\mathbf{\Pi}}(\mathbf{r})\\
&= \left( k^{2} + \partial_{z}^{2} \right) \mathrm{Im}\Pi_{z}.\label{imGfromPi}
\end{align}
Extracting only the $zz$ component of $\mathrm{Im} \paraG(\mathbf{r},\mathbf{r}',\omega)$ is sufficient for our purposes as none of the examples in this work utilise the off-diagonal $z$- components of $\paraG$. However, if needed (for specific dipole orientations in the case of cooperative decay), they can be easily obtained by using appropriate differential operators on $\overline{\mathbf{\Pi}}(\mathbf{r})$. The procedure becomes quite involved when one gets to obtaining the $\overline{\mathbf{\Pi}}(\mathbf{r})$ itself, which requires further elaboration. The technique which we use closely follows the work in \cite{GentiliPelosiPECWedge}; this paper draws inspiration from the vast library of plane-wave diffraction problems treating scattering from wedges (e.g. see \cite{Osipov} and references therein) and adapting them to the case of dipole scattering. We can summarise the procedure: firstly, the fields are represented in terms of plane waves, and secondly, the boundary value problem is solved through the reduction of spectral amplitudes via difference equations. Our aim in this paper is not to explore the detailed methods for finding scattered fields but rather to incorporate some of those ideas into the realm of atomic interactions; we will provide a brief recap of the main steps involved, but we will focus our attention to specific applications concerned with atomic decay rates. 

The main idea relies on a powerful tool employed in finding solutions to scattering problems from a wedge, known as the Sommerfeld integral \cite{Sommerfeld1896}, which we will briefly review. We will start by noting that 
 the solution to the Helmholtz equation for a wedge with an outer angle  $>\pi/2$ is unsuitable for the method of images as adding an `image' wave would not guarantee the solution being free of incoming waves at infinity, therefore violating the radiation condition. The image approach can be rescued by switching an ordinary plane wave solution with a period of $2\pi$ to one with a period of $4\pi$. This transforms the complex plane into a Riemann surface \cite{chew1995waves,donaldson2011riemann} (with the location of the branch points dependent on the particular function), comprising of two sheets, where only one of them permits the existence of incoming waves. This firmly positions the problem in the realm of complex analysis and benefits from its powerful techniques. 
 
 For illustrative purposes, we follow the original reasoning by Sommerfeld \cite{sommerfeld1950lectures},  and tackle a simplified problem in which we seek a function $U(\rho,\phi)$, satisfying a two dimensional --- see the right-hand side of Fig.~\ref{fig:wedge_coords} --- Helmholtz equation. In this case, the source is a plane wave propagating in the plane perpendicular to the $z$- axis. We note that our solution in circular cylindrical coordinates will inevitably involve some plane wave, say $u$, with a constant amplitude $A$, travelling in some direction $\beta$,
 \begin{align}
u = A e^{\mathrm{i}k_{\rho}\rho \cos(\beta-\phi) },
 \end{align}
in which $\mathbf{k}_{\rho} = \hat{x} k_{\rho} \cos \beta + \hat{y}k_{\rho} \sin \beta$, where $k_{\rho} = \sqrt{k_{x}^{2} + k_{y}^{2}}$, and $\boldsymbol{\rho} = \hat{x}\rho \cos \phi + \hat{y}\rho \sin \phi$, such that $\mathbf{k}_{\rho}\cdot \boldsymbol{\rho} = k_{\rho}\rho\cos(\beta-\phi)$. We can assume that a solution satisfying some particular boundary conditions can be expressed as a superposition (bundle) of plane waves. To this end, we employ an infinite summation of Bessel functions --- itself related to plane waves via Sommerfeld's identity \cite{chew1995waves} --- of order $\nu$,
 \begin{align}
 \label{USumNu}
U(\rho,\phi) = \sum_{\nu =0}^{\infty}a_{\nu}(\phi)J_{\nu}(k_{\rho}\rho),
 \end{align}
 weighted by some function $a_\nu(\phi)$. Now, using an integral representation for $J_{n}(x)$ \cite{morse1953methods},
 \begin{align}
 \label{Jofn}
J_{n}(x) = \frac{1}{2\pi}e^{-\mathrm{i}n\pi/2}\int_{C} \meas{\alpha}~e^{\mathrm{i}x \cos\alpha + \mathrm{i}n\alpha},
 \end{align}
we write a solution $U(\rho,\phi)$, in terms of a spectral integral over a contour $C$, which now contains a function of incidence angle/frequency,  $S(\beta)$, as
 \begin{equation}
U(\rho,\phi) = \frac{1}{2\pi \mathrm{i}}\int_{C} \meas{\alpha}~S(\alpha)e^{\mathrm{i}k_{\rho}\rho \cos(\alpha-\phi)}.
 \end{equation}
Such a formulation expressing a solution to the Helmholtz equation as a spectral continuum of plane waves is known as a Sommerfeld integral, and the function $S(\beta)$ that satisfies wedge boundary conditions is a sum (difference) of cotangent functions \cite{Bowman1970}, resulting from evaluating the infinite sum appearing in Eq.~(\ref{USumNu}), with the help of Eq.~(\ref{Jofn}). The path $C$ can be defined to run along the contour $\gamma = \gamma_{\mathrm{L}}+\gamma_{-} +\gamma_{\mathrm{R}} + \gamma_{+}$, seen in Fig.~\ref{complex} and known as a Sommerfeld-Maliuzhinets contour.  This contour has been closed despite the presence of periodically occurring branch points; additionally, the ends of the contour --- situated in the non-shaded regions --- ensure that the $U(\rho,\phi)\rightarrow0$ as $\mathrm{Im}\beta\rightarrow\pm\infty$. Since the dashed contours in Fig.~\ref{complex}, run in opposite directions, they cancel out, i.e. $\gamma_{\mathrm{L}} + \gamma_{\mathrm{R}}=0$, leaving us with
\begin{align}
U(\rho,\phi) &= \frac{1}{2\pi i}\left( \int_{\gamma_{+}} + \int_{\gamma_{-}}\right)  \meas{\alpha'}~ S(\alpha')e^{\i k_{\rho} \rho \cos (\phi-\alpha') }\\
\label{symmetrized}
&= \frac{1}{2\pi i} \int_{\gamma_{+}} \meas{\alpha} \left[ S(\alpha + \phi)-S(-\alpha + \phi)\right]e^{\i k_{\rho} \rho \cos \alpha},
\end{align}
due to the odd parity of $S(\alpha + \phi)$, and the symmetry of Sommerfeld-Maliuzhinets path.

\begin{figure}
    \includegraphics[width=1\linewidth]{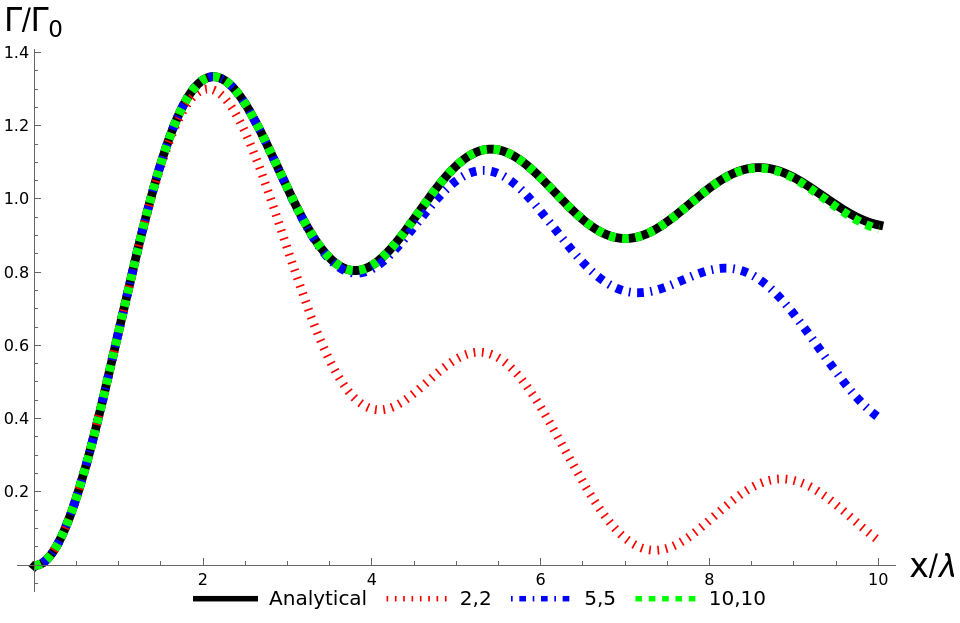}
    \caption{Convergence of the decay rate $\Gamma/\Gamma_{0}$, for a parallel oriented atom, expressed as the truncated summations with $M$, $N$ terms, in comparison to the analytical solution for the case of a semi-infinite plane (wedge with angle $\omega_{0}=\pi$).}
    \label{fig:conv}
\end{figure}

Expanding on this approach, we now concentrate on the problem of the scattering from the wedge in the presence of a $z$- oriented dipole. For this case, the solution will coincide with that of a point source multiplied by $\hat{z}$ \cite{Bowman1970},
\begin{equation}
\label{BowmanPi}
\overline{\boldsymbol{\Pi}}(\mathbf{r})\cdot \hat{z} = U(\rho,\phi,z),
\end{equation}
where
\begin{align}
U(\rho,\phi,z) = \frac{1}{16 \mathrm{i}\pi^{2} \nu}\int\limits_{\gamma_{+}+\gamma_{-}}\meas{\alpha}~\frac{e^{\mathrm{i}kR(\alpha)}}{k R(\alpha)} S(\alpha+\phi), 
\end{align}
in which
\begin{align}
\label{Skernel}
S(\alpha+\phi)=\cot\dfrac{\pi-\alpha-\phi+\phi'}{2\nu} -\cot\dfrac{\pi-\alpha-\phi-\phi'+ \Omega}{2\nu},
\end{align}
and
\begin{equation}
\nu \pi = 2\pi - 2\Omega = \omega_{0},
\end{equation}
while the distance function is
\begin{equation}
R(\alpha)= \sqrt{\rho^{2}+\rho'^{2}+2\rho\rho_{0}\cos\alpha+(z-z')^{2}}.
\end{equation}
We take inspiration from \cite{GentiliPelosiPECWedge}, where a general solution for an arbitrary dipole is sought in the form of an integral over a single loop (due to the function in Eq.~\eqref{Skernel} being odd), $\gamma_{+}$ of the Sommerfeld-Maliuzhinets contour \cite{Mal58,MalTuz62} depicted in Fig.~\ref{complex}; for a comprehensive review of Sommerfeld (and Maliuzhinets) integrals we refer the interested reader to the modern review in Ref.~\cite{Osipov}, as well as Chapter 6 of \cite{Bowman1970}. The contour $\gamma$, has branch points at $\alpha = \pm 2\pi m\pm i \xi$, where $m\in\mathbb{N}$, and $\xi = \cosh^{-1}{\left[ \rho^{2}+\rho'^{2}+(z-z')^{2}/(2\rho \rho')\right]}$, shown as black circles in Fig.~\ref{complex}.    
We thus follow \cite{GentiliPelosiPECWedge} and cast the solution via Eq.~\eqref{BowmanPi} as
\begin{align}
\label{Mal}
\Pi_{z}(\mathbf{r}) = -\frac{1}{2\pi i}\int_{\gamma_{+}}\frac{e^{-i k R(\alpha)}}{R(\alpha)}\pi_{z}\left( \alpha + \frac{n \pi}{2} - \phi \right) 
\meas{\beta},
\end{align}
where we have rewritten the kernel function from Eq.~(\ref{Skernel}) in a symmetrical form as  
\begin{align}
\pi_{z}\Big( \alpha + \frac{\nu \pi}{2} & - \phi \Big) = \frac{1}{8 \pi \varepsilon k Z q} \notag \\
\times \Bigg[ &\cot\left( \frac{\alpha- \Delta^{-}}{2 \nu}  \right) 
- \cot\left( \frac{\alpha-\Delta^{+}}{2 \nu}\right)\notag \\
+ &\cot\left( \frac{\alpha+ \Delta^{-}}{2 \nu}  \right)
- \cot\left( \frac{\alpha+ \Delta^{+}}{2 \nu} \right) \Bigg],
\end{align}
in which we defined $\Delta^{\pm}=\phi\pm\phi'$. By expressing the cotangents in terms of a Fourier series,
\begin{equation}
\cot x = \mathrm{i} -2\mathrm{i}\sum_{k}e^{2k\mathrm{i}x},   
\end{equation} and converting the integral over the radial wave $\int_{\gamma_{+}}\meas{\alpha}\frac{e^{-\mathrm{i}kR(\alpha)}}{R(\alpha)}$, to a series of Hankel functions (\cite{GentiliPelosiPECWedge,Bowman1970}), we can finally write the imaginary $z$- component of the Hertz vector via Eq.~(\ref{Mal}), as
\begin{widetext}
\begin{equation}
\label{PiZ}
 \mathrm{Im}\Pi_{z} = \frac{1}{2 \pi \varepsilon k Z \nu}\sqrt{\frac{2 k \pi}{R_{1}}}\sum_{m=0}^{\infty}\left\{ \sin\left( \frac{m}{\nu}\phi'\right)\sin\left( \frac{m}{\nu}\phi\right) \sum_{p=0}^{\infty}\left[ \frac{J_{\frac{m}{\nu} + 2p +\frac{1}{2}}(k R_{1})}{p! \Gamma(\frac{m}{\nu} + p + 1)} \left( k \frac{\rho \rho'}{2 R_{1}}\right)^{\frac{m}{\nu}+ 2 p} \right] \right\},   
\end{equation}
\end{widetext}
where $R_{1} = \sqrt{\rho^{2}+\rho'^{2}+(z-z')^{2}}$ , and $J_{\nu}(x)$ is a Bessel function of order $\nu$ (a consequence of taking the imaginary part of $i H_{\nu}^{(i)}(x) = J_{\nu}(x)$, for a Hankel function of the type $i$ and order $\nu$). While (\ref{PiZ}) looks formidable, it is easily computed as it depends only on discrete summations rather than (highly-oscillatory) integrals, giving $\mathrm{Im}\paraG_{zz}(\mathbf{r},\mathbf{r},\omega)$ via Eq.~(\ref{imGfromPi}). Since the imaginary part of $\mathbb{G}(\mathbf{r},\mathbf{r},\omega)$ is free of singularities at the coincidence limit, it converges very quickly, and --- as can be seen in Fig.~\ref{fig:conv} for a number of terms $M=N =10$  ---  it allows us to obtain a very close approximation to the analytical result (for distances up to $10 \lambda$ in the specific case of a semi-infinite plane). 
\subsection{Green's tensor components originating from a perpendicular dipole}
In this section, we seek solutions for the remaining parts of the entire Green's tensor, that is, $\perpG(\mathbf{r},\mathbf{r}',\omega)$. We adopt the approach used by Tai \cite{tai1994dyadic}, where the Green's dyadics are expressed in terms of cylindrical vector wave functions. In the following, we will retrace the steps outlined by Buyukdura and Goad \cite{Buyukdura1996}, whose approach modifies Tai's formulation to include spherical wave functions and associated Legendre functions, allowing one to express the solutions not as integrals but as (once more) doubly-infinite series, only this time as expansions of spherical vector functions in terms of spherical coordinates $\mathbf{r} = (r, \theta, \phi)$. 

Even though we require only an intermediate step in finding the electric field, we will briefly outline the strategy to find $\mathbf{E}(\mathbf{r})$, and then specialise to obtain the object of interest, i.e. $\mathrm{Im}\perpG(\mathbf{r},\mathbf{r}',\omega)$.  
Following \cite{Buyukdura1996}, we begin with scalar wave functions $\psi(\mathbf{r})$, solving
\begin{align}
\left(\nabla^{2} + k^{2} \right)\psi(\mathbf{r}) = 0.
\end{align}
The scalar functions are the foundation for constructing the vector wave counterparts, ultimately leading to the spherical wave expansion of the Green's tensor. To satisfy the boundary condition at the surface of our perfect conductor, two types of scalar wave functions are required: at the interface, the odd functions $\psi_{o}(\mathbf{r})$ satisfy the Dirichlet condition
\begin{align}
\label{scalar_helm}
\psi_{o}(\mathbf{r}) = 0,
\end{align}
and the even functions $\psi_{e}(\mathbf{r})$ fulfil the Neumann condition
\begin{align}
\frac{\partial}{\partial n} \psi_{e}(\mathbf{r}) = \frac{\partial}{\partial \phi}\psi_{e}(\mathbf{r}) = 0,
\end{align}
where $n$ is the direction of the unit vector $\hat{n}$ defined to be normal to the wedge's surface, pointing into the (vacuum) background.

The radiation condition, along with the requirement that the functions be finite at $r=0$, $\theta = 0$ and $\theta = \pi$, allows one to solve Eq.~\eqref{scalar_helm} by separation of variables and find the appropriate even and odd wave functions; these are \cite{Buyukdura1996}
\begin{equation}
\label{psi_one}
 \psi_{\mu n \substack{e\\o} }^{(i)} = z^{(i)}_{\mu+n}(k r) T_{\mu+n}^{-\mu}(\cos\theta) \genfrac{}{}{0pt}{}{\cos\mu\phi }{\sin\mu\phi} 
\end{equation}
where $n\in \mathbb{N}$ and
\begin{align}
z_{\nu}^{(1)}(x) =& j_{\nu}(x)\\
z_{\nu}^{(2)}(x) =& h_{\nu}^{(2)}(x)
\end{align}
are spherical Bessel functions and spherical Hankel functions of the second kind, respectively, and $T_{\nu}^{-\lambda}(x)$ is an associated Legendre polynomial. The index $\mu$ is dependent on the wedge angle $\omega_{0}$,
\begin{align}
    \mu = \frac{m\pi}{\omega_{0}},
\end{align}
where $m\in \mathbb{N}$.
We can now define the complete orthogonal sets of vector wave functions: proportional to $j_{\nu}$ and $h_{\nu}^{(2)}$ respectively, given by 
\begin{align}
\overline{\mathbf{M}}^{(i)}_{\mu n}(k \mathbf{r}) &= \nabla \times [k \mathbf{r}\psi_{e}^{(i)}(k \mathbf{r})]\\
\overline{\mathbf{N}}^{(i)}_{\mu n}(k \mathbf{r}) &= \nabla\times\nabla \times [k \mathbf{r}\psi_{o}^{(i)}(k \mathbf{r})].
\end{align}
The above can be re-expressed in terms of auxiliary functions that are independent of $r$
\begin{align}
   \overline{\mathbf{M}}^{(i)}_{\mu n}(k \mathbf{r}) &=  k z^{(i)}_{\mu+n}(k r) \overline{\mathbf{m}}_{\mu n}(\theta,\phi), \\
   \overline{\mathbf{N}}^{(i)}_{\mu n}(k \mathbf{r}) &= \frac{1}{r} z^{(i)}_{\mu+n}(k r) \overline{\mathbf{l}}_{\mu n}(\theta,\phi)\notag \\
   &+ \dfrac{\frac{\mathrm{d}}{\mathrm{d}r}[r z^{(i)}_{\mu + n}(k r)]}{r} \overline{\mathbf{n}}_{\mu n}(\theta,\phi),
\end{align}
where
\begin{align}
 \overline{\mathbf{m}}_{\mu n}(\theta,\phi) &= -\frac{\mu \sin \phi T_{\mu+ n}^{-\mu}(\cos\theta)}{\sin\theta}\hat{\theta}\notag \\
 &-\cos\mu\phi \frac{\mathrm{d}}{\mathrm{d}\theta}[T_{\mu + n}^{-\mu}(\cos\theta)]\hat{\phi},\\
 \overline{\mathbf{n}}_{\mu n}(\theta,\phi) &= \sin\mu\theta \frac{\mathrm{d}}{\mathrm{d}\theta}[T_{\mu+n}^{-\mu}(\cos\theta)]\hat{\theta}\notag\\
 &+ \frac{\mu \cos\mu\theta T_{\mu+n}^{-\mu}(\cos\theta)}{\sin\theta}\hat{\phi},
 \end{align}
 while
\begin{equation}
\overline{\mathbf{l}}_{\mu n}(\theta,\phi) = (\mu + n)(\mu+n+1)\sin\mu\phi T^{-\mu}_{\mu+n}(\cos\theta)\hat{r}.
\end{equation}

Finally, the Green's tensor, expressed with the help of the vector wave functions, reads \cite{Buyukdura1996}
\begin{widetext}
    \begin{equation}
    \label{G tensor sums}
\mathbb{G}(\mathbf{r},\mathbf{r}',\omega) = \frac{\hat{r}\otimes\hat{r}}{k^{2}}\delta(\mathbf{r}-\mathbf{r}') + \left(\frac{\mathrm{i}\pi}{2 k} \right) 
\left\{
\begin{array}{ll}
\displaystyle \sum_{m=0}^{\infty}\sum_{n=0}^{\infty}\frac{\overline{\mathbf{M}}^{(2)}_{\mu n}(k \mathbf{r})\otimes \overline{\mathbf{M}}^{(1)}_{\mu n}(k\mathbf{r}'  ) + \overline{\mathbf{N}}^{(2)}_{\mu n}(k \mathbf{r})\otimes \overline{\mathbf{N}}^{(1)}_{\mu n}(k \mathbf{r}')}{(\mu + n)(\mu + n + 1)Z_{\mu n}} ,&  r >  r' \\
\displaystyle \sum_{m=0}^{\infty}\sum_{n=0}^{\infty}\frac{\overline{\mathbf{M}}^{(1)}_{\mu n}(k \mathbf{r})\otimes \overline{\mathbf{M}}^{(2)}_{\mu n}(k\mathbf{r}'  ) + \overline{\mathbf{N}}^{(1)}_{\mu n}(k \mathbf{r})\otimes \overline{\mathbf{N}}^{(2)}_{\mu n}(k \mathbf{r}')}{(\mu + n)(\mu + n + 1)Z_{\mu n}} ,&  r <  r',
\end{array}
\right.
\end{equation}
\end{widetext}
where 
\begin{align}
Z_{\mu n } = \frac{(1+\delta_{m 0})\pi \omega_{0} n!}{2(2\mu + 2 n + 1)\Gamma(2\mu + n + 1)}.    
\end{align}
The above equation contains a doubly-infinite series, which is divergent at the point $\mathbf{r}'\rightarrow \mathbf{r}$ due to the delta function term and discontinuous at the source's surface $r = r'$. Despite that, the original paper by Buyukdura and Goad \cite{Buyukdura1996} considers Eq.~\eqref{G tensor sums} to be meaningful in situations where the observation point is located far away from the source, and for these cases demonstrate a point-wise convergence. For us, since we are interested in $\mathbb{G}$ at the coincidence limit ($\mathbf{r}\rightarrow \mathbf{r}'$), this seems, on the surface, problematic. However, as was the case with the ``parallel'' part $\paraG$, because we require only the imaginary part of $\mathbb{G}$, the issue of convergence and discontinuity can be resolved. This can be seen by first looking at the definition of the spherical Hankel function (of the second kind)
$
h_{\nu}^{(2)}(x) = j_{\nu}(x) - \mathrm{i}y_{\nu}(x)$,
where $y_{\nu}(x)$ is itself a Bessel function of the second kind. The pairs of the Bessel functions appear in the tensor components in Eq.~(\ref{G tensor sums}) as terms proportional to $\mathrm{i}j_{\nu}(k r)(j_{\nu}(kr')-\mathrm{i}y_{\nu}(kr'))$, and $\mathrm{i}j_{\nu}(k r')(j_{\nu}(kr)-\mathrm{i}y_{\nu}(kr))$, thus extracting the imaginary components in each case leaves us with the product $j_{\nu}(kr)j_{\nu}(kr')$. This results in the $r>r'$ and $r<r'$ cases being equal, making the imaginary part of the Green's tensor continuous at $r = r'$. In addition, discarding the (real) delta function term by considering only the imaginary part of $\mathbb{G}$ completes solving the issues of divergence and discontinuity. We can now write our second main result, the imaginary Green's tensor expanded in terms of the spherical vector wave functions, which, after taking the coincidence limit, is given by     
\begin{widetext}
\begin{equation}
\mathrm{Im}\mathbb{G}(\mathbf{r},\mathbf{r},\omega) = \frac{\pi}{2 k}\sum_{m=0}^{\infty}\sum_{n=0}^{\infty}\frac{\overline{\mathbf{M}}^{(1)}_{\mu n}(k \mathbf{r})\otimes \overline{\mathbf{M}}^{(1)}_{\mu n}(k\mathbf{r}  ) + \overline{\mathbf{N}}^{(1)}_{\mu n}(k \mathbf{r})\otimes \overline{\mathbf{N}}^{(1)}_{\mu n}(k \mathbf{r})}{(\mu + n)(\mu + n + 1)Z_{\mu n}}.
\label{ImGperp}
\end{equation}
\end{widetext}
The resulting (imaginary) Green's tensor can be easily transformed into a Cartesian basis, allowing for picking out the appropriate matrix components, using the prescription for $\perpG(\mathbf{r},\mathbf{r},\omega)$ given by Eq.~(\ref{G perp}). In a similar spirit to Eq.~(\ref{PiZ}), this complicated formula can be --- for our purposes ---  computed accurately with only a small number of terms (a cutoff of ten in each sum suffices).  Due to the spherical nature of the expansion functions, calculating the term $\mathbb{G}_{zz}(\mathbf{r},\mathbf{r},\omega)$ (dipole oriented in $z$) is not preferable; evaluating it accurately would require computing significantly more terms in the expansions, making the process inefficient. Hence, calculating fields from a parallel ($z$ oriented) dipole is better accomplished using the cylindrical wave expansion presented in the previous section.

\section{Examples and possible applications}
\label{sec5}
\subsection{Decay Rate}

\begin{figure*}

        \includegraphics[width=0.45\textwidth]{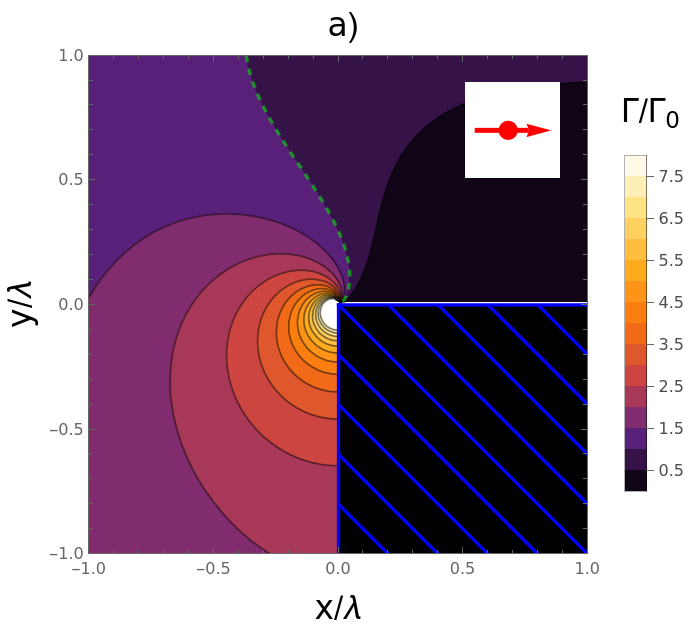}
        \includegraphics[width=0.45\textwidth]{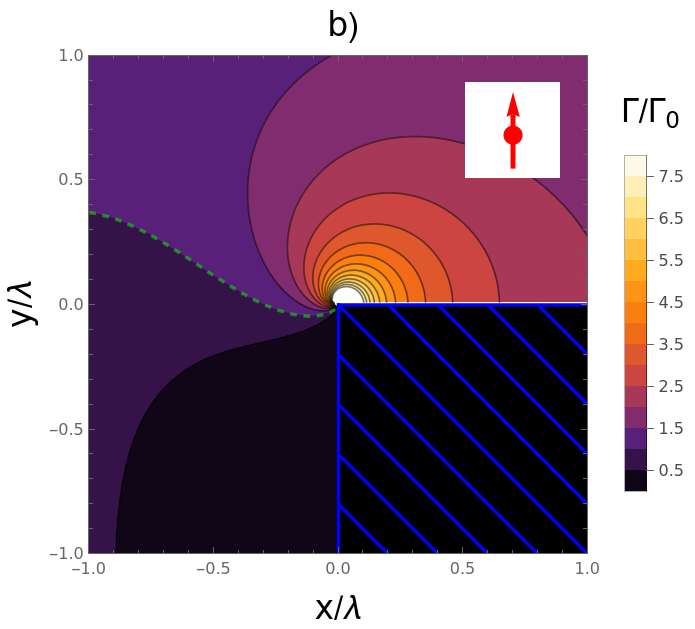}
        \includegraphics[width=0.45\textwidth]{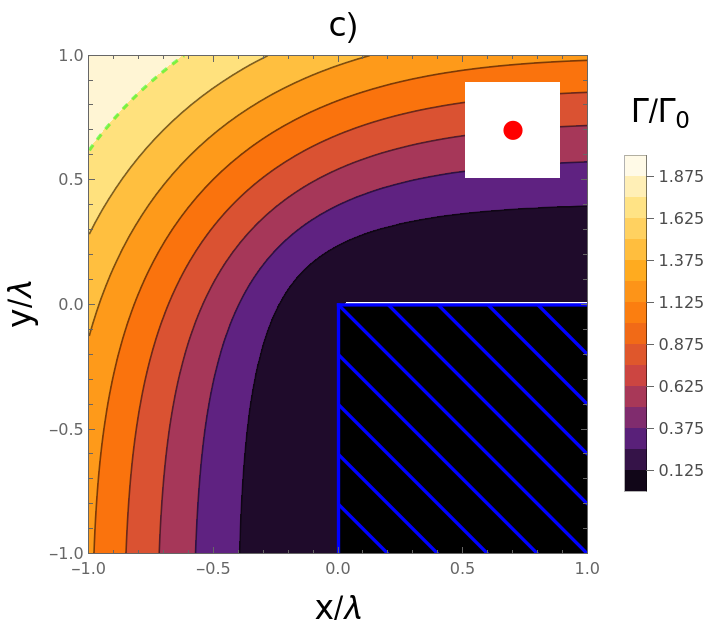}
        \includegraphics[width=0.45\textwidth]{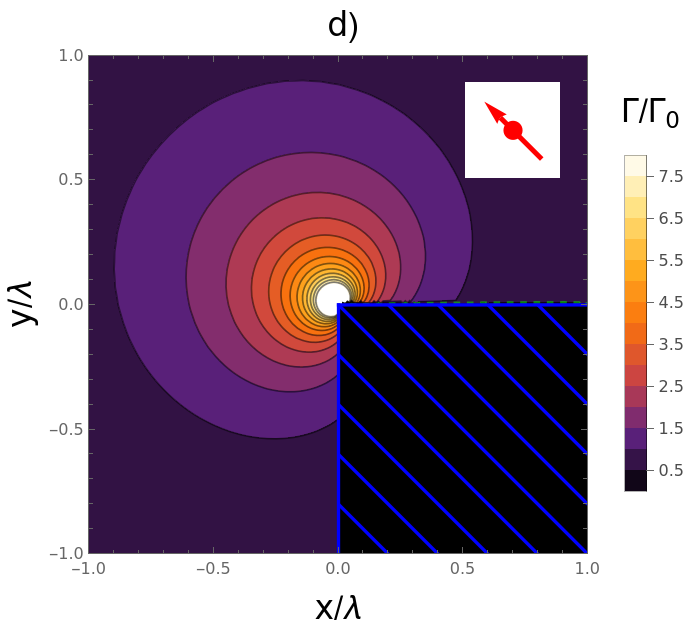}
    \caption{Normalised decay rate $\Gamma/\Gamma_{0}$ in vacuum near a wedge-shaped perfect electric conductor (with $\Gamma_{0}$ being the decay rate in vacuum). Each coordinate represents the decay rate at the coincidence limit, $\mathbf{r}'\rightarrow \mathbf{r}$. The cartoon of the dipole (in red) in the top right corner shows the orientation of the dipole moment with respect to the PEC surface (blue lines on the black background). The green dashed line shows the contour where the normalised decay rate equals unity --- or, in other words, where the modified rate is the same as the vacuum rate.}
    \label{fig:decayRate}
\end{figure*}

The normalised spontaneous decay rate as a function of the position is depicted in Fig.~\ref{fig:decayRate}, where we used the wedge angle $\omega_{0}=\pi/2$ to showcase the two extremes of the effects connected to the dipole orientation.
As expected, we observe a symmetric behaviour between the plots in Fig.~\ref{fig:decayRate}a and Fig.~\ref{fig:decayRate}b, as the ``bright''  side of the wedge for the $y-$ oriented dipole becomes the ``dark'' side of the wedge for the $x-$ oriented dipole.  Flipping the dipole's orientation switches the regions of suppression and enhancement, as expected. At the distances $\lesssim \lambda/2$, we start to see a significant enhancement of the decay rate on the ``bright'' side, which continues to rise as we approach the corner --- the sudden change in the relative dipole orientation with respect to the closest side of the wedge creates an effect of a discontinuity, which in turn, acts like an antenna,  enhancing the LDOS for the perpendicular orientation of the dipole. In contrast, in the case of the (always) parallel dipole, shown in Fig.~\ref{fig:decayRate}c, the discontinuity does not form at all, and we observe a smooth decay rate, approximating that of a semi-infinite plane ``bent around'' the corner. As suspected, for the dipole oriented along the diagonal in the $x$-$y$ plane (at a $135^{\circ}$ angle to the $x$-axis) as shown in Fig.~\ref{fig:decayRate}d, we see a symmetrical decay rate, the enhancement occurring on either side of the plate due to the relative point of discontinuity being symmetrical in this case. Tracing the decay rate along the line of symmetry, we see that it qualitatively follows the decay rate of a perpendicularly polarised dipole for the semi-infinite plate, albeit with an enhanced rate near the origin.  

\begin{figure}
    \includegraphics[width=1\linewidth]{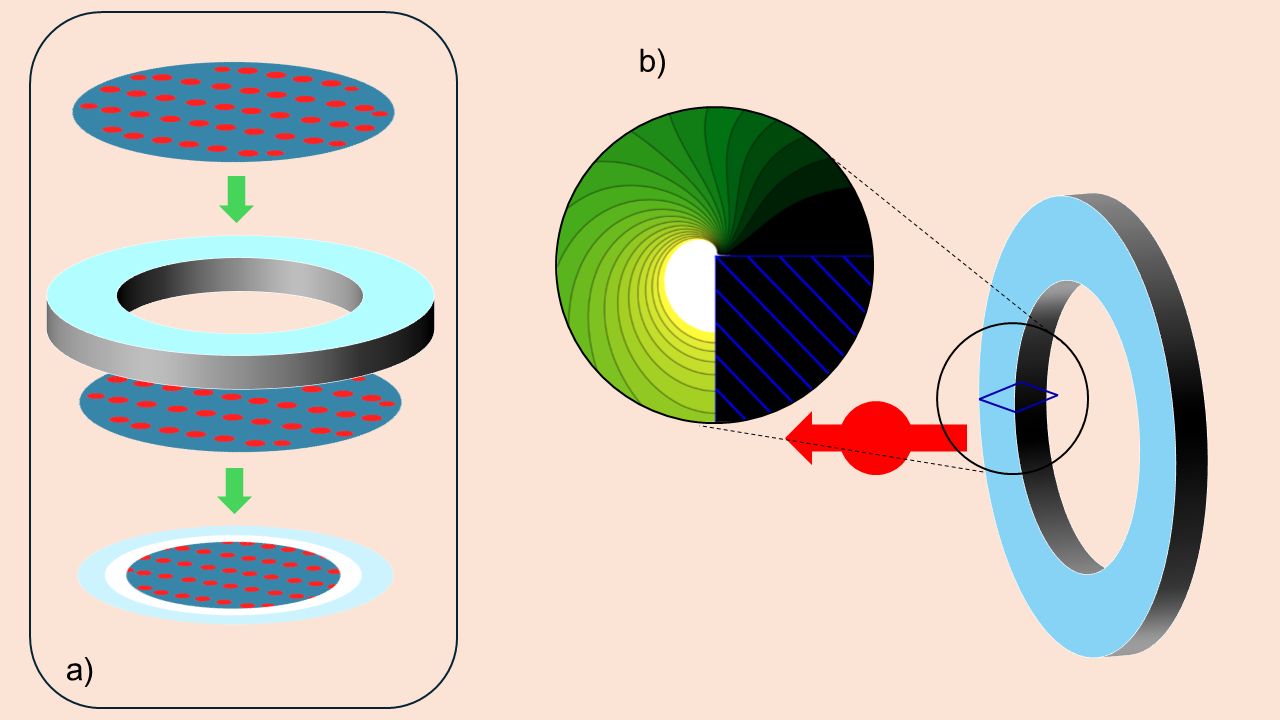}
    \caption{Schematic of operation of a metallic ring-based quenching for possible use in microscopy. a) shows a cartoon of the process - a sample filled with ready-to-fluoresce molecules interacts with a metallic ring. As a result, the sample's spontaneous emission is altered, and molecules in the affected regions (light blue corresponding to the cross-section of the plate and white to the highly enhanced regions of the hole) decay faster, reducing the focal spot. b) shows the metallic plate with an opening reacting to the atomic dipole (in red), where we indicated the emphasis on the corner element- showing a zoomed-in example of the spontaneous decay rate for a perpendicular-oriented atom.  }
    \label{fig:STED}
\end{figure}

The most interesting effects occur in the vicinity of the corner; looking at Fig.~\ref{fig:decayRate}a and Fig.~\ref{fig:decayRate}b, we examine the behaviour of the dipole near the edge of the metallic wedge --- the situation represented in the right panel of Fig.~\ref{fig:STED}.
As we approach the corner along the side to which the dipole moment is orthogonal, the decay rate increases up to ten times for distances $\sim 0.01 \lambda$ from the corner, compared to the two-fold increase at the face of a half-plate. Conversely, approaching from the other side offers a negligible decay rate enhancement compared to the far-from-the-edge rate. 
For distances far away from the corner its effects on the decay landscape diminish as the solutions approach those of the semi-infinite plane. These intricacies cannot be captured by a simplified approach in which the corner is considered as two perpendicular flat (infinite) plates and the solutions combined ad-hoc, showing that the proper consideration of the effects of the corner is vital for accurately describing the decay rates in its vicinity.

Moving onto potential applications of the modified decay rates, we imagine a possible modification to the STED technique. This would entail removing the need for its second, depleting laser and replacing the effect of stimulated emission with an increased spontaneous emission rate generated by a (sharply-edged) metallic ring. The primary motivation behind this idea is reducing the chance of photobleaching \cite{PBOracz2017} when imaging biological samples and simplifying the process in some respects by removing the complication of aligning and calibrating the second laser. The modified procedure can, therefore, be summarised as follows: after the initial beam has activated the dye molecules --- making them ready to fluoresce --- we introduce a metallic ring at a distance of $\sim 0.1 \lambda$ above the sample, shown in the left panel of Fig.~\ref{fig:STED}. This has the effect of accelerating the spontaneous decay rate (two-fold) at the locations where the ring would ``cast a shadow'' (the light blue outer ring at the bottom cartoon on the left side of Fig.~\ref{fig:STED}), and causing an even more considerable increase due to the corner effects in the neighbouring ring of an approximate thickness of half a wavelength (the white ring at the bottom of left panel in Fig.~\ref{fig:STED}). Now, between the time of application of the metallic ring and the time of unaffected spontaneous decay $\tau_{0} (\approx 5 \text{ns})$, a scan of the region with reduced focal size can be performed, offering an increased contrast relative to the situation with no ring. 
Interestingly, the effect of a high increase in the decay rate stems from purely geometrical characteristics of the system, i.e. the presence of a sharp corner. The geometrical singularities that increase field effects known from everyday life, such as, e.g. a lightning rod, have been explored extensively concerning fluorescence - an interested reader may see some recent works in, e.g. \cite{LiRoHe2021, LiRoUrbieta2018}.

We will now estimate the efficacy of such a technique. In the preceding sections, we analysed the behaviour of molecules whose emission was polarised in different directions with respect to the metallic edge. In the following, we make a simplifying assumption that the molecules are polarised parallel to the ring's symmetry axis. The dipoles orthogonal to it would not experience an azimuthally uniform enhancement of their decay rates; thus, we leave those cases from our analysis.   
    
\begin{figure}
    \centering
    \includegraphics[width=1\linewidth]{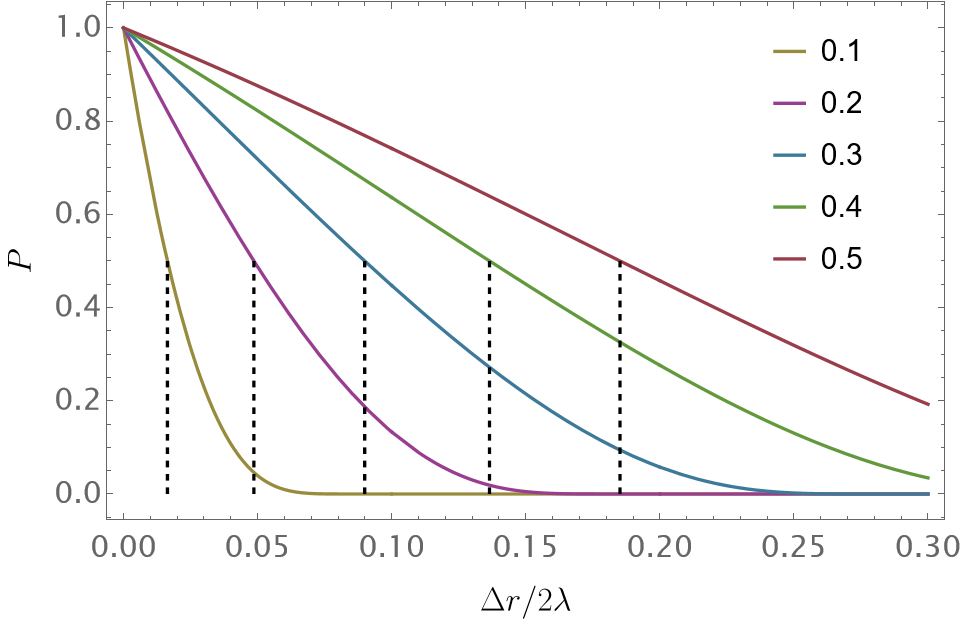}
    \caption{Probability of photon detection curves for different radii $R$, of the hole in the metallic ring, and locations of (half-) spot sizes - marked as dashed, vertical lines.}
    \label{fig:FWHM}
\end{figure}

We can perform a loose approximation of the size of the focal spot by calculating the full-width half-maximum (FWHM) of the probability pulse at time $\tau_{0}$, defined as the time when the number of spontaneously decayed states has dropped to a factor $e^{-1}$ of its initial value. We modify Equation ($1$) from Ref.~\cite{STEDHell} to account for adding the metallic mask with radius $R$ and exclude the STED beam. We model the probability to detect a photon at time $\tau_{0}$ --- after the mask has been applied --- as 
\begin{align}
\label{Prob}
P(\mathbf{r};\tau_{0}) = h(\mathbf{r})\eta(\mathbf{r};\tau_{0}),
\end{align}
where $\eta(\mathbf{r};\tau_{0}) = \exp{[-\Gamma(\mathbf{r})\tau_{0}}]$ ($\tau_{0}$ is the inverse of the vacuum decay rate, i.e. $\tau_{0}=1/\Gamma_{0}$), is the normalised decay rate in our geometry, and given a lens with semi aperture $\alpha$,
\begin{equation}
\label{h}
h(\mathbf{r}) = C \cos^{2}{\left[\frac{\pi}{\lambda} (r-R) n \sin{\alpha}\right]}e^{-(r-R)^{2}/2 R^{2}},
\end{equation}
where $C$ is a normalisation constant and $n$ is the refractive index Ref.~\cite{STEDHell}. Eq.~(\ref{h}) represents a normalised beam intensity multiplied by a Gaussian, which modifies the intensity profile to be negligible beyond a distance $R$ --- the hole's radius. We solve Eq.~(\ref{Prob}) for values of $r$ for which the probability $P(\mathbf{r};\tau_{0})$ drops to half of its maximum value. The resulting distance, $\Delta r/2$, gives us the lateral spot size, allowing us to establish the level of improvement beyond the diffraction limit. We have plotted probability curves using $n \sin\alpha \approx 1$, for different radii of the hole and calculated the corresponding $\Delta r/\lambda$ (spot size) - shown in Fig.~\ref{fig:FWHM}.
We see that, unsurprisingly, the smallest $R$ results in the smallest spot size; for $R = 0.1 \lambda$, $\Delta r/2\approx 0.03 \lambda$ and with enlarging the $R$ in increments of $0.1 \lambda$, we notice that the spot size (half), grows linearly by $0.1 \lambda$. Thus, taking the green light as an example, $\lambda \approx 500 \text{nm}$, an opening of $R = 50 \text{nm}$, we could theoretically achieve a resolution of $\approx 20 \text{nm}$ - comparable to the improvement achievable via conventional STED \cite{STEDHell}.

\subsection{Cooperative Decay Rate}
In this section, we explore a related quantity --- the \textit{cooperative} decay rate (CDR) for which a second dipole modifies the decay landscape in addition to local effects from a sharp, metallic corner. In our investigations of the modified STED, we conceptualised the corner as a limiting shape of a larger structure - a metallic ring. In contrast, we treat the corner more conceptually here. We imagine that the effect of a measurable decay rate affected by an object hidden from a line of view - being ``behind the corner'' - to be of considerable interest \cite{AroundCornersPRL, AroundCornersRapp2020}.  

\begin{figure*}
        \includegraphics[width=0.4\textwidth]{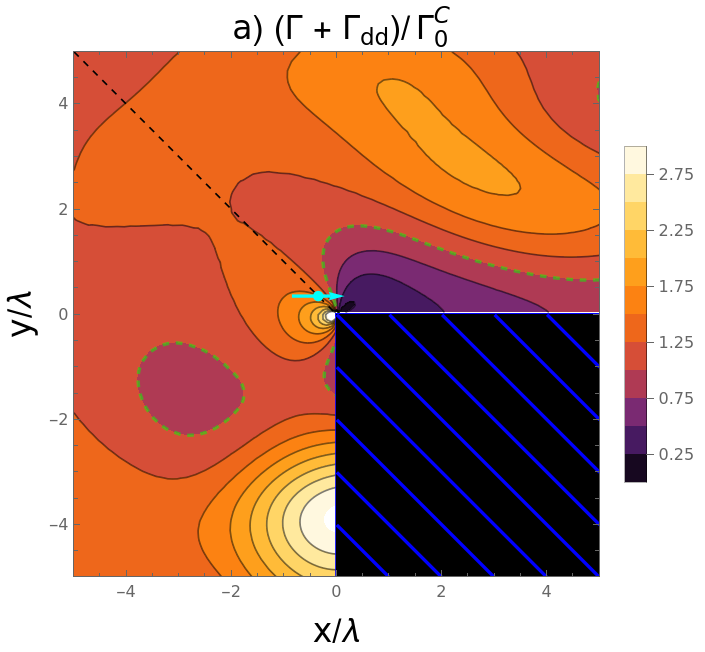}
        \includegraphics[width=0.4\textwidth]{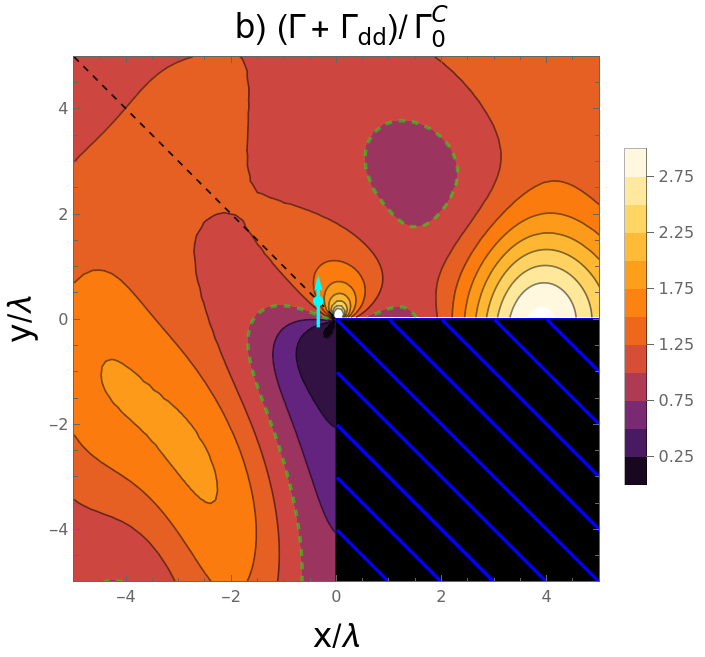}
        \includegraphics[width=0.4\textwidth]{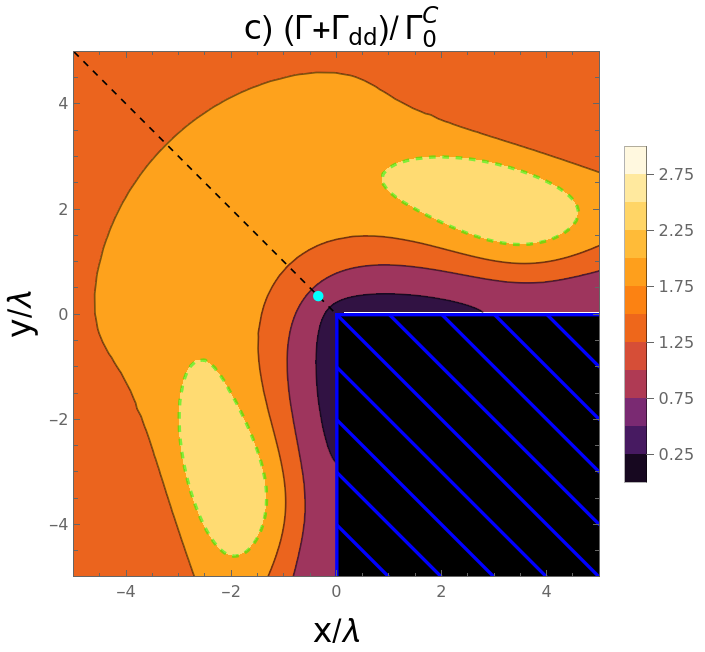}
        \includegraphics[width=0.4\textwidth]{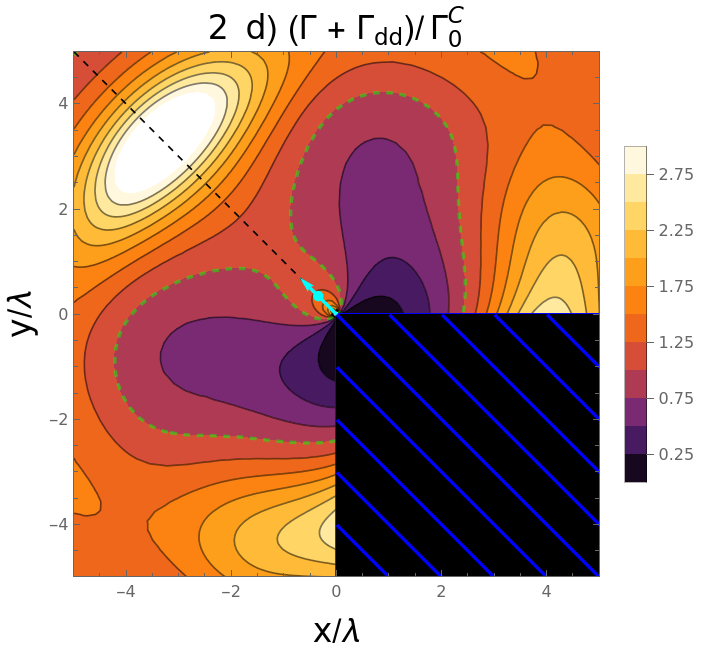}

    \caption{Normalised cooperative decay rate $(\Gamma + \Gamma_{\mathrm{dd}})/\Gamma^{C}_{0}$ in a vacuum, where $\Gamma^{C}_{0}$ is the geometry-free normalisation, in the vicinity of a perfectly reflecting corner. The cartoon of the dipole (in cyan) overlaid on plots (a)-(d) shows the location and the orientation of the donor atom's dipole moment, located at $\mathbf{r}_{\mathrm{D}}=(-c, c,0)$ for $c = \sqrt{2}\lambda/4$, with respect to the conducting surface (blue lines on the black background). The green dashed line shows the contour where the CDR is equal to one - its vacuum rate.}
    \label{fig:CDRplots}
\end{figure*}

We examine the behaviour of two entangled atoms modelled as dipole emitters, where one is excited, and the other is in the ground state, which we label respectively as a donor and an acceptor. The position of the donor atom is fixed on the diagonal symmetry line at a distance of one wavelength away from the corner, and the position of the acceptor atom is variable, as depicted in Fig.~\ref{fig:CDRplots}. For dipoles oriented along 
$\hat{x} $ and $\hat{y}$ respectively, Fig.~\ref{fig:CDRplots}a) and Fig.~\ref{fig:CDRplots}b), we again see a behaviour exhibiting symmetry. Each dipole experiences the highest enhancement of its CDR at a distance of around three wavelengths from the corner on each side. Analogous to the single-atom decay, the ``perpendicular side'' is qualitatively suppressed as we move away from the surface, but the near-the-surface rate is enhanced; the effects on the ``parallel side'' are commensurate in reverse with the ones behind the corner, but as parallel orientation reaches maximum decay away from the surface, the enhancement is not as strong. The always-parallel ($z$-oriented) dipole, shown in Fig.~\ref{fig:CDRplots}c), bears similarities to its single-atom counterpart in Fig.~\ref{fig:decayRate}c), where the dipole orientation aligned with the edge does not experience an enhancement due to the discontinuity, both sides show agreement with the half-space rates shown in Fig.~\ref{fig:EvolPlots}f). The diagonally (at $135^{\circ}$) oriented dipole in the $x$-$y$ produces a decay landscape symmetrical about the diagonal line, where the most significant enhancement occurs not at the tip but approximately three wavelengths away along the diagonal. All the examples are qualitatively similar to the single-atom case as functions of the distance from the face of the wedge (the parallel side is enhancing, and the perpendicular side is suppressing). The difference is seen in displacing the ``epicentre'' of enhancement by the distance of approximately three wavelengths from the tip, true for the non-$z$-oriented dipoles.  
\subsection{Variable wedge angle}

\begin{figure*}
    \includegraphics[width=0.3\textwidth]{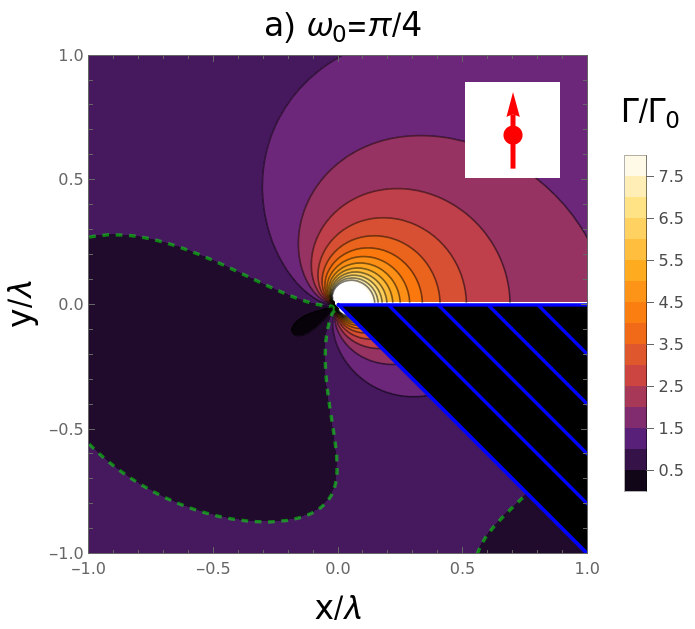}
    \includegraphics[width=0.3\textwidth]{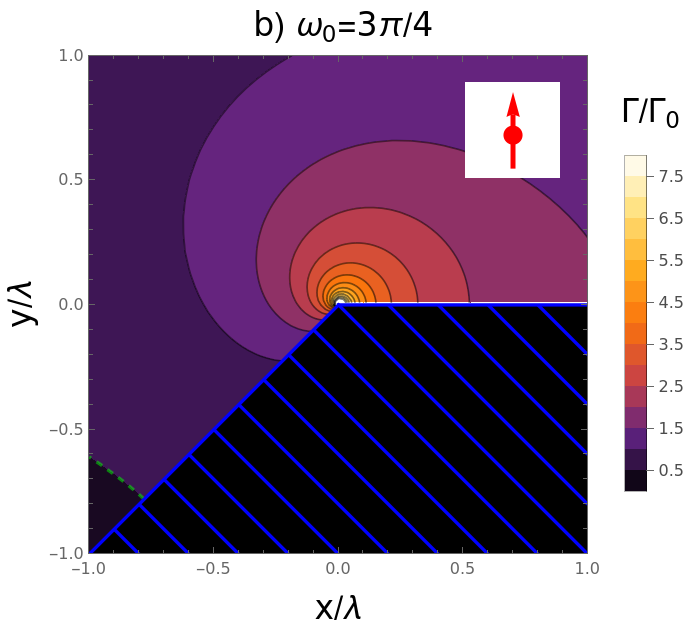}
    \includegraphics[width=0.3\textwidth]{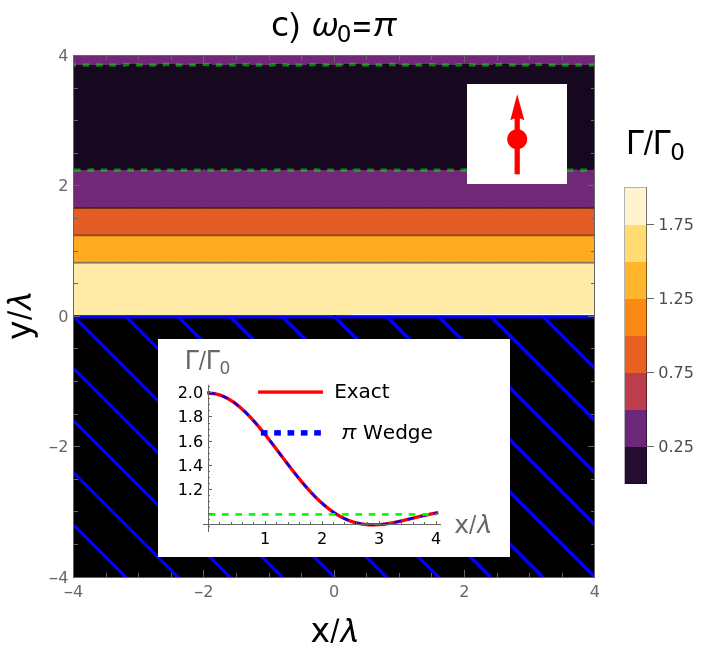}
    \includegraphics[width=0.3\textwidth]{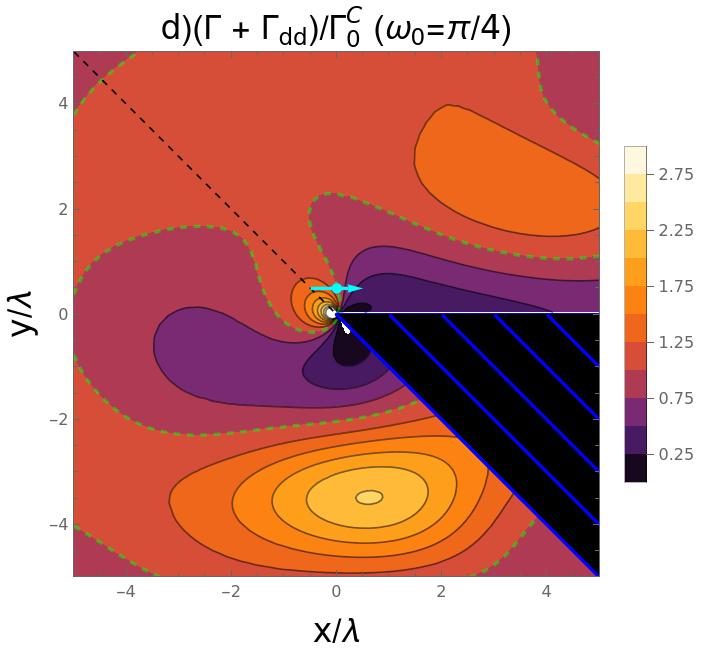}
    \includegraphics[width=0.3\textwidth]{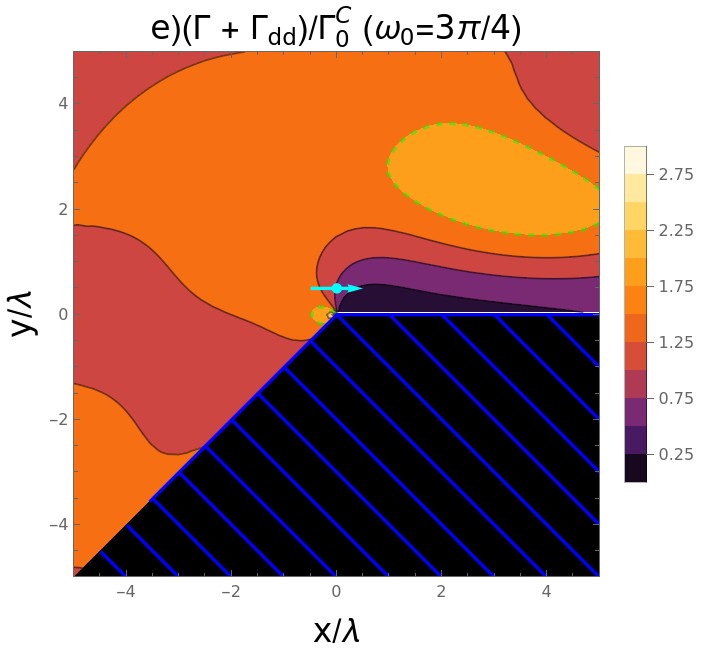}
    \includegraphics[width=0.3\textwidth]{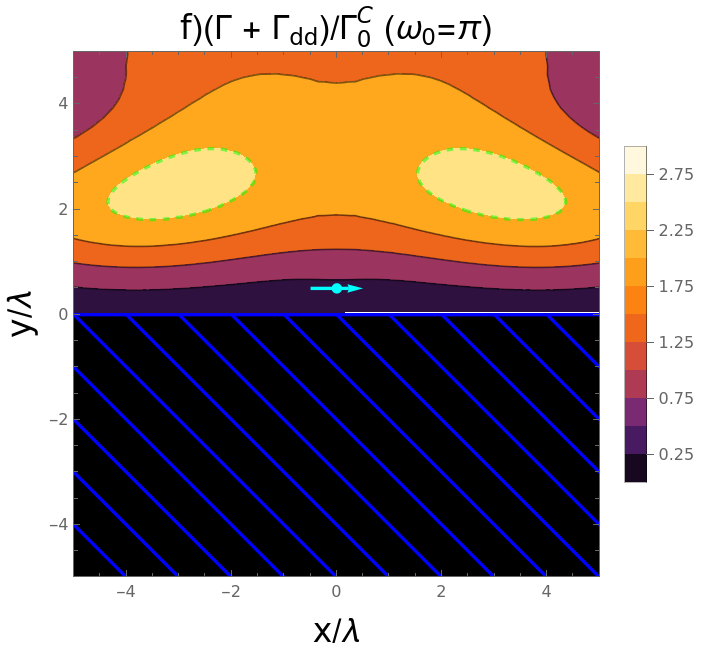}

    \caption{Decay rates for a selection of different wedge angles $\omega_{0}$ of a PEC surface (depicted as blue lines on the black background). The top panel, (a)-(c), shows normalized decay rates $\Gamma/\Gamma_{0}$, for a dipole oriented in $y$ direction. Additionally, panel (c) features an inset showing the agreement between the analytical decay rate for a half-space and the one obtained via our method for a wedge with angle $\omega_{0} = \pi$. The bottom panel, (d)-(f), shows the same selection of angles $\omega_{0}$, but depicts normalized CDR $(\Gamma + \Gamma_{\mathrm{dd}})/\Gamma^{C}_{0}$, in the $x$ direction, where the dipole is located at $\mathbf{r}_{\mathrm{D}}=(0,\lambda/2,0)$, and $\Gamma^{C}_{0}$ is the CDR without geometry present. In all cases, the dashed green line shows the respective decay rate equal to one.}
    \label{fig:EvolPlots}
\end{figure*}

Complementary to the corner example, we examined the cases of a variable wedge angle $\omega_{0}$ for both the decay and cooperative decay rates. These are shown in the top and bottom panels of Fig.~\ref{fig:EvolPlots}. This is of particular interest for the case of CDR, as it shows the possibility of detecting mutual coherence between objects around corners of variable angles. As a consistency check, we also note in Fig.~\ref{fig:EvolPlots}c) that our results for a dipole oriented perpendicularly to a `wedge' with an opening angle of $\pi$ (i.e. a flat plane) are in exact agreement with those found analytically for that configuration (complementary to the parallel dipole results in Fig.~\ref{fig:conv}).

\section{Summary and Conclusions}
We have presented an investigation of atomic decay rates in the presence of a sharp, metallic corner. This work combines the formalism of classical electromagnetic theory with quantum mechanics to study atomic behaviour, applicable to modern experiments \cite{DDExpAbbasirad, DDExpPRL,DDExpXu2024} 

We have outlined two approaches for calculating the electric field from a dipole in the presence of a perfectly reflecting wedge, each suitable for calculating different field components. Both methods were initially developed decades ago in the field of antenna radiation --- here, we demonstrated how these could be modernised and adapted to our needs. To demonstrate the correctness and power of the tools, we investigated two observable quantities. Firstly, we looked at the spontaneous decay rate of an atom near a wedge, speculating in detail about possible applications in STED-like microscopy using nano-scale rings. We also estimated an improvement in resolution compared to the diffraction limit. The second result, concerned with cooperative decay rates, is of more theoretical and fundamental interest. It demonstrates how one atom's decay rate can be influenced by another even though it is around the corner. It is somewhat analogous to pursuing similar imaging techniques in classical optics.

\begin{acknowledgments}
R.K. acknowledges financial support from the UK EPSRC Doctoral Training Programme grant EPSRC/DTP 2020/21/EP/T517896/1, and R.B. acknowledges financial support from the UK EPSRC grant  EP/W016486/1. 
\end{acknowledgments}

\appendix

\section{Derivation of the cooperative decay rate}
\label{ap}
\begin{widetext}
We want to derive the cooperative decay rate 
\begin{equation}
\Gamma^{\mathrm{C}} =  \frac{2\omega_{a}^{2}}{\epsilon_{0}c^{2}}\mathbf{d}_{\mathrm{A}}\cdot\left[ \frac{1}{2}\mathrm{Im}\mathbb{G}(\mathbf{r}_{\mathrm{A}},\mathbf{r}_{\mathrm{A}},\omega_{a}) + \frac{1}{2}\mathrm{Im}\mathbb{G}(\mathbf{r}_{\mathrm{D}},\mathbf{r}_{\mathrm{D}},\omega_{a})\pm\mathrm{Im}\mathbb{G}(\mathbf{r}_{\mathrm{A}},\mathbf{r}_{\mathrm{D}},\omega_{a})\right]\cdot\mathbf{d}_{\mathrm{D}} ,
\end{equation}
where $\mathbf{d}_{\mathrm{A}}$ and $\mathbf{d}_{\mathrm{D}}$ are the dipole moments of the atoms positioned respectively at $\mathbf{r}_{\mathrm{A}}$ and $\mathbf{r}_{\mathrm{D}}$, radiating at frequency $\omega \equiv \omega_{\mathrm{AD}}$, and $\mathbb{G}(\mathbf{r}_{\mathrm{A}},\mathbf{r}_{\mathrm{D}},\omega)$ is the Green's tensor evaluated at their respective positions $\mathbf{r}_{\mathrm{A}}$ and $\mathbf{r}_{\mathrm{D}}$.\\

We start from the expression for Fermi's Golden Rule, which describes transition rates from a set of initial states $\ket{i}$ to a set of final states $\ket{f}$
\begin{align}
	\Gamma = \frac{2 \pi}{\hbar}\sum_{i,f}^{} |\bra{f}\hat{H}_{\mathrm{int}}\ket{i}|^{2}\delta(E_{f}-E_{i}),
\end{align}
where $\hat{H}_{\mathrm{int}}$ is the relevant interaction Hamiltonian, and $E_{i/f}$ are the energies of the initial and final states, respectively. 
 
We will assume that our two atoms are initially entangled. Therefore we specify the initial state as symmetric ($+$) or anti-symmetric ($-$) tensor products of the two-atom system with the field
\begin{align}\label{initialState}
\ket{i} &=\dfrac{1}{\sqrt{2}}(\ket{e_{\mathrm{A}};g_{\mathrm{D}}} \pm \ket{g_{\mathrm{A}};e_{\mathrm{D}}}) \otimes \ket{0} \equiv  \dfrac{1}{\sqrt{2}}(\ket{e_{\mathrm{A}};g_{\mathrm{D}};0} \pm \ket{g_{\mathrm{A}};e_{\mathrm{D}};0})
\end{align}
where $\ket{\mathbf{1}_{\lambda}(\mathbf{r},\omega)}$ is understood to be a state with one excitation of the field-matter system, i.e. the result of applying the creation operator defined in \eqref{EFieldDef} and \eqref{commutator} to the vacuum: $\hat{\mathbf{f}}^{\dagger}_{\lambda}(\mathbf{r},\omega)\ket{0}$,  $\lambda$ denotes whether this is an electric- and magnetic-type excitation. The final state will be that with both atoms in the ground state;
\begin{align}\label{finalState}
\ket{f} &= \ket{g_{\mathrm{A}};g_{\mathrm{D}}}\otimes \ket{\mathbf{1}_{\lambda}(\mathbf{r},\omega)}\equiv  \ket{g_{\mathrm{A}};g_{\mathrm{D}};\mathbf{1}_{\lambda}(\mathbf{r},\omega)}
\end{align}
Using the states \eqref{initialState} and \eqref{finalState} Fermi's Golden rule, we have
\begin{align}
	\label{FGR}
\Gamma = \frac{ \pi}{\hbar}\int \mathrm{d}^{3}r'' \int\mathrm{d}\omega' 
&\bra{g_{\mathrm{A}};g_{\mathrm{D}};\mathbf{1}_{\lambda}(\mathbf{r},\omega)}\hat{H}_{int} \left(\ket{e_{\mathrm{A}};g_{\mathrm{D}};0} \pm \ket{g_{\mathrm{A}};e_{\mathrm{D}};0} \right)\\\cdot
&\left(\bra{e_{\mathrm{A}};g_{\mathrm{D}};0} \pm \bra{g_{\mathrm{A}};e_{\mathrm{D}};0}\right) \hat{H}_{\mathrm{int}} \ket{g_{\mathrm{A}};g_{\mathrm{D}};\mathbf{1}_{\lambda}(\mathbf{r},\omega)}
\delta(E_{f}-E_{i})\nonumber
\end{align}
Our interaction Hamiltonian is that of dipoles coupled to the electric field; 
\begin{equation}\label{interactionHamiltonian}
    \hat{H}_{\mathrm{int}} = -\hat{\mathbf{d}}_{\mathrm{A}}\cdot \hat{\mathbf{E}}(\mathbf{r},t) -\hat{\mathbf{d}}_{\mathrm{D}}\cdot \hat{\mathbf{E}}(\mathbf{r},t) \equiv \hat{H}_{\mathrm{int}}^{\mathrm{A}} + \hat{H}_{\mathrm{int}}^{\mathrm{D}}.
\end{equation}

As in Eq.~\eqref{EFieldDef}, the electric field operator  $\hat{\mathbf{E}}(\mathbf{r},t)$ is expressed as
\begin{align}
\hat{\mathbf{E}}(\mathbf{r},t) = \int_{0}^{\infty}\mathrm{d}\omega\sum_{\lambda=e,m}^{}
\int\mathrm{d}^{3}r' [ \mathbb{G}_{\lambda}(\mathbf{r},\mathbf{r'},\omega)\cdot \hat{\mathbf{f}}_{\lambda}(\mathbf{r'},\omega)e^{-i\omega t} +
\mathbb{G}_{\lambda}^{*}(\mathbf{r},\mathbf{r'},\omega)\cdot \hat{\mathbf{f}}^{\dagger}_{\lambda}(\mathbf{r'},\omega)e^{i\omega t}  ].
\end{align}\\
The contribution of the first term in Eq.~\eqref{interactionHamiltonian} to the first line of the rate in rate shown in Eq.~\eqref{FGR} is therefore:
\begin{align}
	\label{Hint}
\bra{g_{\mathrm{A}};g_{\mathrm{D}}; \mathbf{1}_{\lambda}(\mathbf{r''},\omega')}\hat{H}_{\mathrm{int}}^{\mathrm{A}}&\left(\ket{e_{\mathrm{A}};g_{\mathrm{D}};0} \pm \ket{g_{\mathrm{A}};e_{\mathrm{D}};0} \right) = \bra{g_{\mathrm{A}};g_{\mathrm{D}}}\otimes \bra{0}\hat{\mathbf{f}}_\lambda(\mathbf{r''},\omega')~
\hat{\mathbf{d}}_{\mathrm{A}}\cdot \hat{\mathbf{E}}(\mathbf{r}_{\mathrm{A}},t) \ket{e_{\mathrm{A}};g_{\mathrm{D}}} \otimes\ket{0}\notag \\
=& \bra{g_{\mathrm{A}}}\hat{\mathbf{d}}_{\mathrm{A}}\ket{e_{\mathrm{A}}}\nonumber
\int_{0}^{\infty}\mathrm{d}\omega\sum_{\lambda'=e,m}^{}
\int\mathrm{d}^{3}\mathbf{r'}~
\mathbb{G}_{\lambda'}^{*}(\mathbf{r}_{\mathrm{A}},\mathbf{r'},\omega)\bra{0}
\hat{\mathbf{f}}_{\lambda}(\mathbf{r''},\omega') \hat{\mathbf{f}}^{\dagger}_{\lambda'}(\mathbf{r'},\omega)\ket{0}e^{i\omega t} \\
=&~ {\mathbf{d}}_{\mathrm{A}}^{\downarrow}\cdot\int_{0}^{\infty}\mathrm{d}\omega\sum_{\lambda'=e,m}^{}
\int\mathrm{d}^{3}\mathbf{r'}~\mathbb{G}_{\lambda'}^{*}(\mathbf{r}_{\mathrm{A}},\mathbf{r'},\omega) e^{i\omega t}\delta_{\lambda \lambda'}\delta(\omega-\omega')\delta(\mathbf{r'}-\mathbf{r''}) \nonumber \\
=&~{\mathbf{d}}_{\mathrm{A}}^{\downarrow}\cdot \mathbb{G}_{\lambda}^{*}(\mathbf{r}_{\mathrm{A}},\mathbf{r''},\omega)\nonumber e^{i\omega' t},
\end{align}
where we defined ${\mathbf{d}}_{\mathrm{A}}^{\downarrow}\equiv \bra{g_{\mathrm{A}}}\hat{\mathbf{d}}_{\mathrm{A}}\ket{e_{\mathrm{A}}}$ as the dipole matrix element for the `downwards' transition, and we made use of the commutation relation \eqref{commutator}. The term with a $\pm$ disappears due to the orthogonality of atomic states. 

The contribution of the second term in Eq.~\eqref{interactionHamiltonian} to the first line of the rate in rate shown in Eq.~\eqref{FGR} can be calculated in an exactly analogous way, giving:
\begin{align*}
 \bra{g_{\mathrm{A}};g_{\mathrm{D}}; \mathbf{1}_{\lambda}(\mathbf{r''},\omega')}\hat{H}_{\mathrm{int}}^{\mathrm{D}}(\ket{g_{\mathrm{A}};e_{\mathrm{D}};0}\pm\ket{e_{\mathrm{A}};g_{\mathrm{D}};0})  =&\pm\hat{\mathbf{d}}_{\mathrm{D}}^{\downarrow}\cdot \mathbb{G}^{*}(\mathbf{r}_{\mathrm{D}},\mathbf{r''},\omega) e^{i\omega' t},
\end{align*}
where ${\mathbf{d}}_{\mathrm{D}}^{\downarrow}\equiv \bra{g_{\mathrm{A}}}\hat{\mathbf{d}}_{\mathrm{D}}\ket{e_{\mathrm{A}}}$. Similarly, we obtain the contributions of both terms in Eq.~\eqref{interactionHamiltonian} to the second line of the rate in rate shown in Eq.~\eqref{FGR} as:
\begin{align}
\left(\bra{e_{\mathrm{A}};g_{\mathrm{D}};0} \pm \bra{g_{\mathrm{A}};e_{\mathrm{D}};0}\right) \hat{H}_{\mathrm{int}}^{\mathrm{A}} \ket{g_{\mathrm{A}};g_{\mathrm{D}};\mathbf{1}_{\lambda}(\mathbf{r},\omega)} &= ~ \hat{\mathbf{d}}_{\mathrm{A}}^{\uparrow}\cdot \mathbb{G}(\mathbf{r}_{\mathrm{A}},\mathbf{r''},\omega) e^{-i\omega' t} + 0\\
\left(\bra{e_{\mathrm{A}};g_{\mathrm{D}};0} \pm \bra{g_{\mathrm{A}};e_{\mathrm{D}};0}\right) \hat{H}_{\mathrm{int}}^{\mathrm{D}} \ket{g_{\mathrm{A}};g_{\mathrm{D}};\mathbf{1}_{\lambda}(\mathbf{r},\omega)} &= \pm \hat{\mathbf{d}}_{\mathrm{D}}^{\uparrow}\cdot \mathbb{G}(\mathbf{r}_{\mathrm{D}},\mathbf{r''},\omega) e^{-i\omega' t} + 0.
\end{align}
where ${\mathbf{d}}_{\mathrm{D/A}}^{\uparrow}\equiv \bra{e_{\mathrm{D/A}}}{\mathbf{d}}_{\mathrm{D/A}}\ket{g_{\mathrm{D/A}}}$. Putting everything together,  we arrive at 
\begin{equation}
\begin{aligned}
|\bra{f}\hat{H}_{\mathrm{int}}\ket{i}|^{2} = \frac{1}{2}\left[ {\mathbf{d}}_{\mathrm{A}}^{\downarrow}\cdot\mathbb{G}^{*}(\mathbf{r}_{\mathrm{A}},\mathbf{r''},\omega) \pm {\mathbf{d}}_{\mathrm{D}}^{\downarrow} \cdot \mathbb{G}^{*}(\mathbf{r}_{\mathrm{D}},\mathbf{r''},\omega) \right] \cdot
\left[ {\mathbf{d}}_{\mathrm{A}}^{\uparrow}\cdot\mathbb{G}(\mathbf{r}_{\mathrm{A}},\mathbf{r''},\omega) \pm  {\mathbf{d}}_{\mathrm{D}}^{\uparrow} \cdot \mathbb{G}(\mathbf{r}_{\mathrm{D}},\mathbf{r''},\omega) \right] ,
\end{aligned}
\end{equation}\\
which can be simplified using the fact that for vectors $\mathbf{u},\mathbf{v}$ and matrices $\mathbb{M},\mathbb{N}$, the identity
$(\mathbf{u\cdot\mathbb{M}})\cdot (\mathbf{v}\cdot\mathbb{N}) = \mathbf{u}\cdot (\mathbb{M}\cdot \mathbb{N}^{\mathrm{T}})\cdot \mathbf{v}$, holds, giving us
\begin{align}
	\label{FourExp}
2|\bra{f}\hat{H}_{\mathrm{int}}\ket{i}|^{2} &= \left[ {\mathbf{d}}_{\mathrm{A}}^{\downarrow}\cdot
\mathbb{G}_{\lambda'}(\mathbf{r}_{\mathrm{A}},\mathbf{r''},\omega') ~\mathbb{G^{*}}_{\lambda'}^{\mathrm{T}}(\mathbf{r}_{\mathrm{A}},\mathbf{r''},\omega')
\cdot {\mathbf{d}}_{\mathrm{A}}^{\uparrow}\right]\notag \\
&\quad +\left[ {\mathbf{d}}_{\mathrm{A}}^{\downarrow}\cdot
\mathbb{G}_{\lambda'}(\mathbf{r}_{\mathrm{A}},\mathbf{r''},\omega') ~\mathbb{G^{*}}_{\lambda'}^{\mathrm{T}}(\mathbf{r}_{\mathrm{D}},\mathbf{r''},\omega')
\cdot {\mathbf{d}}_{\mathrm{D}}^{\uparrow}
 \right]\notag \\
 &\quad\quad+\left[ {\mathbf{d}}_{\mathrm{D}}^{\downarrow}\cdot
 \mathbb{G}_{\lambda'}(\mathbf{r}_{\mathrm{D}},\mathbf{r''},\omega') ~\mathbb{G^{*}}_{\lambda'}^{\mathrm{T}}(\mathbf{r}_{\mathrm{A}},\mathbf{r''},\omega')
 \cdot {\mathbf{d}}_{\mathrm{A}}^{\uparrow}
 \right]\notag \\
  &\quad\quad\quad+\left[ {\mathbf{d}}_{\mathrm{D}}^{\downarrow}\cdot
 \mathbb{G}_{\lambda'}(\mathbf{r}_{\mathrm{D}},\mathbf{r''},\omega') ~\mathbb{G^{*}}_{\lambda'}^{\mathrm{T}}(\mathbf{r}_{\mathrm{D}},\mathbf{r''},\omega')
 \cdot {\mathbf{d}}_{\mathrm{D}}^{\uparrow}
 \right].
\end{align}
Using the integral relation \cite{BuhmannPRA}
\begin{align}
\sum_{\lambda=e,m}^{}\int \mathrm{d}^{3}\mathbf{s}~ \mathbb{G}_{\lambda}(\mathbf{r},\mathbf{s},\omega)\cdot \mathbb{G^{*}}_{\lambda}(\mathbf{r'},\mathbf{s},\omega) = \frac{\hbar\mu_{0}}{\pi}\omega^{2}\mathrm{Im}~ \mathbb{G}(\mathbf{r},\mathbf{r'},\omega)
\end{align}
we can now insert the expression in Eq.~(\ref{FourExp}) into Fermi's Golden rule. Assuming that the dipoles are oriented in the same direction and possess the same magnitude i.e. ${\mathbf{d}}_{\mathrm{D}}^{\downarrow} = {\mathbf{d}}_{\mathrm{D}}^{\uparrow} =  {\mathbf{d}}_{\mathrm{A}}^{\downarrow} = {\mathbf{d}}_{\mathrm{A}}^{\uparrow} = |\mathbf{d}|\mathbf{\hat{n}} $, where $\mathbf{\hat{n}}$ is a unit vector, we obtain
\begin{align}
\Gamma^{\mathrm{C}} = 2\frac{\pi \hbar \mu_{0}|\mathbf{d}|^{2}}{\hbar\pi}\int \mathrm{d}\omega' \omega'^{2}\mathbf{\hat{n}}\cdot\left[\frac{1}{2}\mathrm{Im}\mathbb{G}(\mathbf{r}_{\mathrm{A}},\mathbf{r}_{\mathrm{A}},\omega') +
\frac{1}{2}\mathrm{Im}\mathbb{G}(\mathbf{r}_{\mathrm{D}},\mathbf{r}_{\mathrm{D}},\omega') \pm \mathrm{Im}\mathbb{G}(\mathbf{r}_{\mathrm{A}},\mathbf{r}_{\mathrm{D}},\omega')\right]\cdot \mathbf{\hat{n}}\delta(E_{f} -E_{i}),
\end{align}
here $E_{f} -E_{i} = \hbar \omega' - \hbar \omega_{a}$, $\hbar \omega_{a}$ being the atomic transition frequency. Integrating the expression above, we obtain
\begin{align}
\label{CDR final}
\Gamma^{\mathrm{C}} =  \frac{2|\mathbf{d}|^{2}\omega_{a}^{2}}{\epsilon_{0}c^{2}}\mathbf{\hat{n}}\cdot\left[ \frac{1}{2}\mathrm{Im}\mathbb{G}(\mathbf{r}_{\mathrm{A}},\mathbf{r}_{A},\omega_{a}) + \frac{1}{2}\mathrm{Im}\mathbb{G}(\mathbf{r}_{\mathrm{D}},\mathbf{r}_{\mathrm{D}},\omega_{a})\pm\mathrm{Im}\mathbb{G}(\mathbf{r}_{\mathrm{A}},\mathbf{r}_{\mathrm{D}},\omega_{a})\right]\cdot\mathbf{\hat{n}} ,
\end{align}
which we can write as
\begin{align}
\Gamma^{\mathrm{C}} = \Gamma \pm \Gamma_{\mathrm{dd}},
\end{align}
 either atom ($\mathrm{A}$ or $\mathrm{D}$)  can spontaneously decay to the ground state, and we call this transition $\Gamma$; the additional term $\Gamma_{\mathrm{dd}}$ is the dipole-dipole interaction coupling which for positive sign in front results in superradiance, and a suppressed rate (subradiance) in the opposite case. The normalised CDR is computed by plugging in the analytical form of the vacuum Green's tensor $\mathbb{G}=\mathbb{G}^{(0)}$ into (\ref{CDR final}), being mindful of the fact that at the coincidence, $\mathbb{G}^{(0)}(\mathbf{r},\mathbf{r},\omega)$ is a diagonal matrix.
\end{widetext}

\bibliography{CDR}

\end{document}